%% file: main.tex
\documentclass[acmsmall]{acmart}

\usepackage{tabularx}
\usepackage{booktabs}
\usepackage{soul}
\AtBeginDocument{%
  \providecommand\BibTeX{{%
    \normalfont B\kern-0.5em{\scshape i\kern-0.25em b}\kern-0.8em\TeX}}}

\setcopyright{none}
\copyrightyear{}
\acmYear{}
\acmDOI{}

\acmPrice{}
\acmISBN{}
\acmArticleSeq{0}

\begin{document}

\title[Human-Generative-AI Interactions]{An HCI-Centric Survey and Taxonomy of Human-Generative-AI Interactions}


\author{Jingyu Shi}
\authornote{Both authors contributed equally to this research.}
\orcid{0000-0001-5159-2235}
\email{shi537@purdue.edu}
\affiliation{%
  \institution{Department of Electrical and Computer Engineering, Purdue University}
  \city{West Lafayette}
  \state{Indiana}
  \country{USA}
}

\author{Rahul Jain}
\authornotemark[1]
\email{jain348@purdue.edu}
\orcid{0009-0001-3723-5482}
\affiliation{%
  \institution{Department of Electrical and Computer Engineering, Purdue University}
  \city{West Lafayette}
  \state{Indiana}
  \country{USA}
}

\author{Hyungjun Doh}
\orcid{0009-0008-3154-1201}
\email{hdoh@purdue.edu}
\affiliation{%
  \institution{Department of Electrical and Computer Engineering, Purdue University}
  \city{West Lafayette}
  \state{Indiana}
  \country{USA}
}

\author{Ryo Suzuki}
\orcid{0000-0003-3294-9555}
\email{ryo.suzuki@ucalgary.ca}
\affiliation{%
  \institution{Department of Computer Science, University of Calgary}
  \city{Calgary}
  \state{Alberta}
  \country{Canada}
}

\author{Karthik Ramani}
\orcid{0000-0001-8639-5135}
\email{ramani@purdue.edu}
\affiliation{%
  \institution{Department of Mechanical Engineering, Purdue University}
  \city{West Lafayette}
  \state{Indiana}
  \country{USA}
}

\thanks{This work is partially supported by the NSF under the Future of Work at the Human-Technology Frontier (FW-HTF) 1839971. We also acknowledge the Feddersen Distinguished Professorship Funds and a gift from Thomas J. Malott. Any opinions, findings, and conclusions expressed in this material are those of the authors and do not necessarily reflect the views of the funding agency.}
\authorsaddresses{Authors’ addresses: Jingyu S., Rahul J., Hyungjun D., Karthik R.: West Lafayette, IN, USA. Ryo S.: Calgary, AL, Canada.}

\renewcommand{\shortauthors}{Shi and Jain, et al.}

\input{texes/0-Abstract}


\maketitle
\input{texes/1-Introduction}
\input{texes/2-Background_Scopes}
\input{texes/2-Methodology}

\input{texes/3-Purpose}
\input{texes/4-Feedback}
\input{texes/5-Control}

\input{texes/6-Engagement}
\input{texes/7-Applications}
\input{texes/8-Evaluations}
\input{texes/9-Discussion}
\input{texes/91-Future}

\input{texes/92-Conclusion}


\bibliographystyle{ACM-Reference-Format}
\bibliography{main}

\input{texes/93-Appendix}

\end{document}

%% file: texes/0-Abstract.tex
\begin{abstract}
Generative AI (GenAI) has shown remarkable capabilities in generating diverse and realistic content across different formats.
HCI literature has investigated how to effectively create collaborations between humans and GenAI systems.
However, the current literature lacks a comprehensive framework to better understand human-GenAI interactions, as the holistic aspects of human-centered GenAI systems are rarely analyzed systematically.
We present a survey of 291 papers, providing a novel taxonomy and analysis of Human-GenAI Interactions from both human and GenAI perspectives.
We highlight challenges and opportunities to guide the design of GenAI systems and interactions toward the future design of human-centered GenAI applications.
\end{abstract}

\ccsdesc[500]{Human-centered computing~Interaction design theory, concepts and paradigms}

\keywords{generative AI, human-AI interaction, interactive machine learning}

\received{20 February 2007}
\received[revised]{12 March 2009}
\received[accepted]{5 June 2009}

%% file: texes/1-Introduction.tex
\section{Introduction}
Recently, Generative Artificial Intelligence (GenAI) models have gained immense popularity and are being applied in diverse applications such as art~\cite{li2023scaling, alayrac2022flamingo}, design~\cite{karras2019style, sbai2018design}, and entertainment~\cite{li2021inco, tseng2023edge}.
Current popular GenAI models including Large Language Models (LLM) and Large Vision Models (LVM) are widely deployed on platforms or in software for their capabilities to create imagery content (Dalle-2~\cite{ramesh2022hierarchical}, Stable Diffusion~\cite{rombach2022high}), writing literature~\cite{yuan2022wordcraft}, and Question Answering (ChatGPT~\cite{openai2021chatgpt}).
The adoption of GenAI models has demonstrated significant advantages, e.g. fostering creativity~\cite{qiao2022initial}, driving innovation~\cite{chang2023prompt}, enabling personalized content generation~\cite{wu2023styleme}, and providing valuable assistance in creation endeavors~\cite{ma2022ai}. 
As a result, these models have become ubiquitous, emphasizing the need for well-crafted and compelling interactions between humans and GenAI.

To take advantage of the generative power of GenAI models, Human-GenAI interaction techniques such as prompt engineering~\cite{liu2022design, qiao2022initial}, visualization~\cite{wan2023gancollage}, and interactive interfaces~\cite{dang2022ganslider, evirgen2023ganravel}, have become popular and effective mediums for humans to interact with GenAI systems.
These allow users to collaborate~\cite{yuan2022wordcraft}, be assisted\cite{aksan2018deepwriting}, take suggestions~\cite{chu2023wordgesture} or revise recommendations~\cite{valencia2023less} from GenAI systems.

However, existing research on Human-GenAI interactions focuses on each individual aspect and domain.
The key design considerations, common practices, and future research opportunities are still hidden and scattered across many broad topics embedded in GenAI and its applications.
To keep pace with the development of GenAI models and their new out-of-the-box capabilities, we identify a need to systematically analyze the research in this field, particularly from an interaction design perspective, to assist the HCI community to innovate and explore new interaction design techniques for the best utilization of GenAI capabilities.
Furthermore, this view of GenAI will also foster new emerging applications to consider key vantage points where our framework will point.

Inspired by the above, we aimed to lay the groundwork for further developments in the field of human-GenAI interactions by systematically synthesizing all the current research and consolidating the existing knowledge and approaches in this domain. 

\begin{figure}[htp]
  \includegraphics[width=\textwidth]{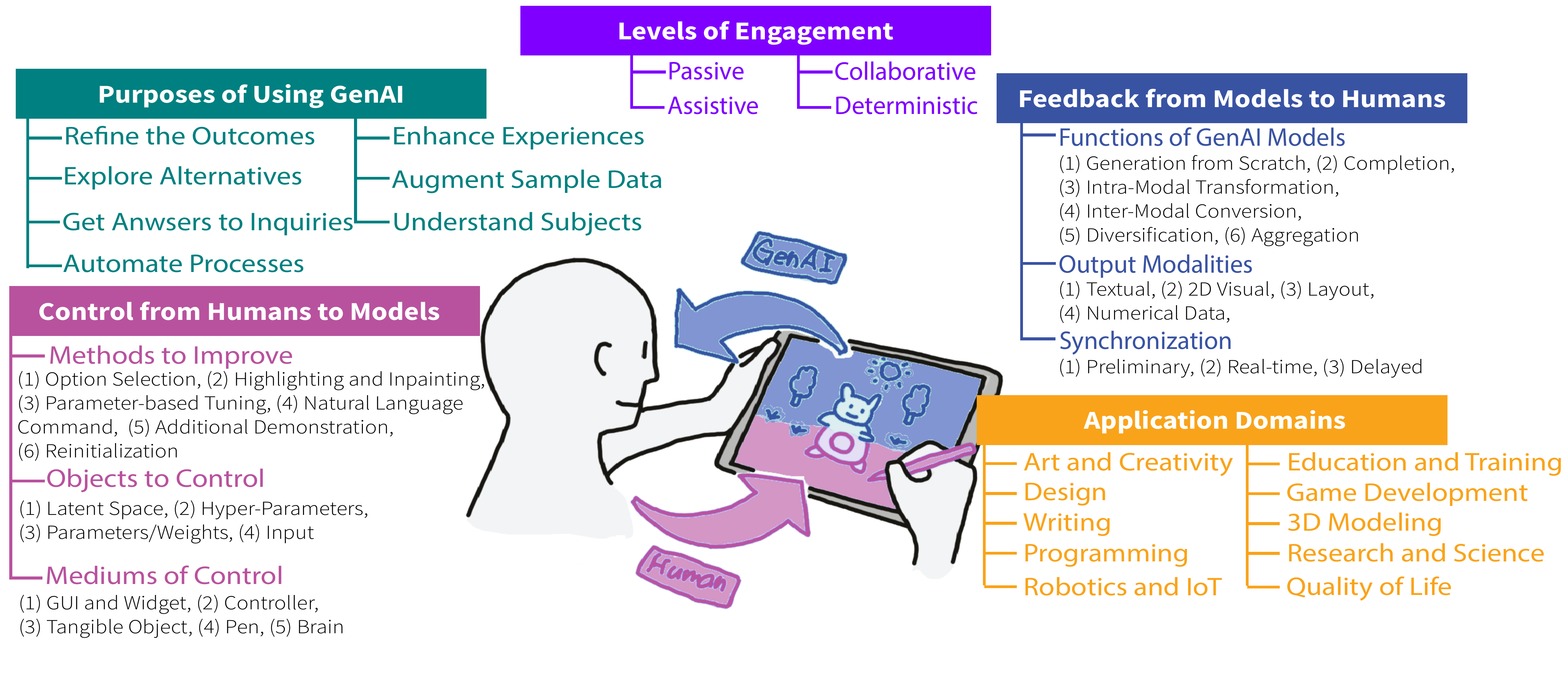}
  \caption{Visual abstract of our survey and taxonomy of Human-GenAI Interaction. Our taxonomy summarizes five key dimensions, namely, Purposes of Using GenAI, Feedback from Models to Humans, Control from Humans to Models, Levels of Engagement, and Application Domains.}
  \label{fig:teaser}
\end{figure}

In this paper, we review a corpus of 291 papers for synthesizing the taxonomy of human-GenAI interactions. 
Specifically, we synthesize the research fields from both the user and GenAI perspectives (briefly shown in ~\autoref{fig:teaser}) into the following dimensions of the design space: 1) Purposes of Using GenAI, 2) Feedback from Models to Users, 3) Control from Users to Models, 4) Levels of Engagement, 5) Application Domains, and 6) Evaluation Strategies.

Our main goal is to present a comprehensive overview of recent developments in and research on AI-model analysis, interaction designs, visualization techniques, and application domains of GenAI-based systems.
By compiling the state-of-the-art advancements in these areas, we aim to provide a valuable resource for researchers to understand the current landscape and situate their own work within a broader design space.
To achieve the goals above, we summarize a taxonomy from the literature, offering a holistic view that encompasses perspectives from both the GenAI model side (e.g. I/O design, capabilities, and volumes) and the human side (e.g. evaluation strategies and application domains), as well as the interactions between them (e.g. interfaces to control, visualization technique, and feedback design). 
This taxonomy will enable readers to gain a deeper understanding of the intricacies involved in creating effective and meaningful human-GenAI interaction systems, fostering future evolution and innovation in the design of GenAI technologies.
Additionally, we identify open research questions, challenges, and opportunities in the future design of GenAI systems and interactions.
By highlighting these areas of exploration, we aim to guide researchers in their pursuit of addressing crucial issues and uncovering new possibilities, so that the HCI community paces and also identifies vantage points for creating new GenAI, with the rapidly evolving technology of GenAI.

%% file: texes/2-Background_Scopes.tex
\section{Background, Scopes, and Contributions}

In this section, we cover the developments in GenAI models in prior research as well as open-source platforms and software. 

\begin{figure*}[ht]
    \centering
    \includegraphics[width=\textwidth]{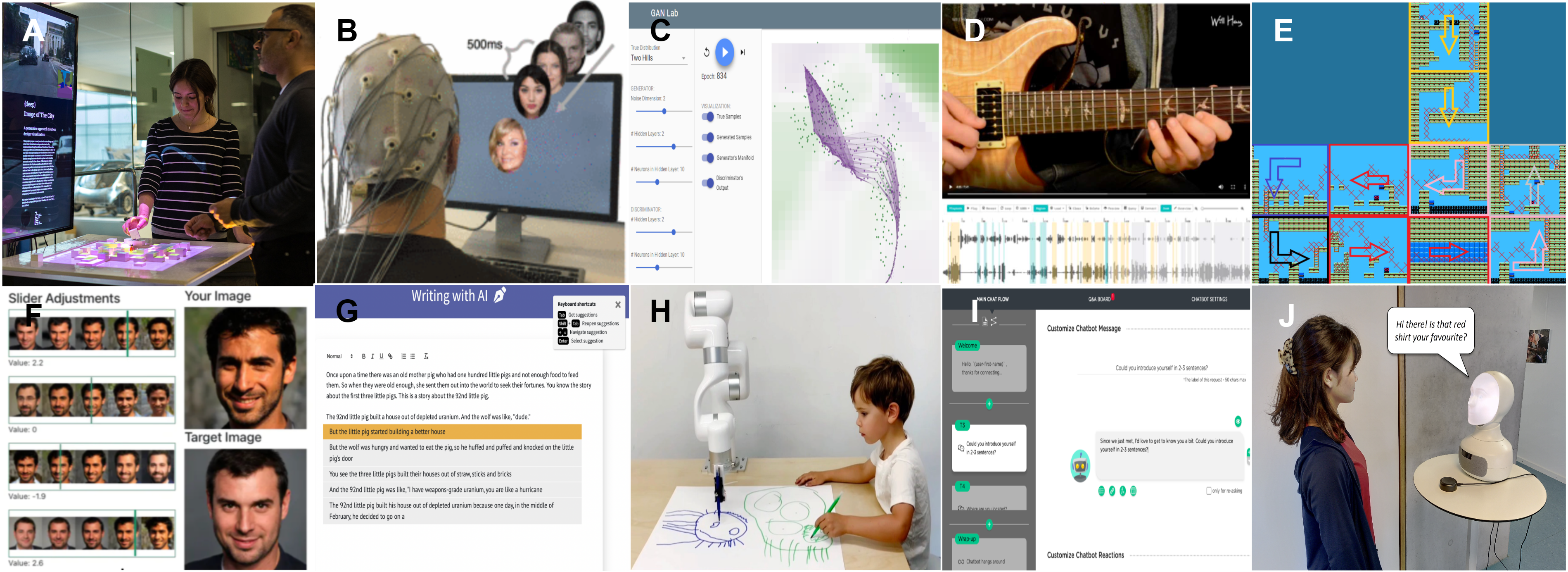}
    \caption{Examples of GenAI applications located in our survey, covering the topics of research: A) Embodied interactions with GenAI~\cite{noyman2020deepscope}, B) Direct Control human to AI~\cite{spape2021brain}, C) Human Interpretable~\cite{kahng2018gan}, D) Gen AI enhancing skill~\cite{wang2021soloist}, E)  Automate Process \cite{capps2021using}, F) Human controllability~\cite{dang2022ganslider}, G) Natural Language Generation \cite{goodman2022lampost}, H) Human AI collaboration~\cite{twomey2022three}, I) Personalization and Adaptation \cite{han2021designing}, J) Conversational GenAI~\cite{janssens2022cool}} 
    \label{fig:examples}
\end{figure*}

A widely accepted definition~\cite{xu2015overview} of GenAI goes by \textit{the probabilistic models that model the joint distribution} in contrast to discriminate AI models \textit{that model the conditional distribution}.
From a high-level understanding, GenAI is defined as "an AI that uses existing media to create new, plausible media"~\cite{muller2022genaichi}.
As a sub-topic in the domain of AI, GenAI has a rich history of development.

In the early stage of its development, the metaphor of GenAI advanced independently in two major domains, namely Natural Language Processing (NLP) and Computer Vision (CV).
In NLP, the generation of nature languages by AI was handled by early implementations of Recurrent Neural Networks(RNN)~\cite{elman1990finding} and Long Short-Term Memory (LSTM) networks~\cite{graves2012long}.
Likewise in CV, the concepts of Artificial/Convolutional Neural Networks (ANN~\cite{fukushima1988neocognitron} and CNN~\cite{lecun1998gradient}) were applied to models that generate images.
Nevertheless, in both fields, the NN-based methods were greatly limited by the hardware conditions back then.

It was not until the early 2010s that the breakthrough in hardware technology enabled the explosion in GenAI research with increased computational power in both NLP and CV fields.
Long sentence generation and sequence-to-sequence generation were then achieved by RNNs with much larger sizes and computational power~\cite{sutskever2014sequence, mikolov2013efficient}.
Similarly in the CV area, models like Generative Adversarial Networks (GAN)~\cite{goodfellow2014generative}, Variational Autoencoders (VAE)~\cite{kingma2013auto}, and their successors~\cite{karras2019style, karras2017progressive, karras2020analyzing, brock2018large} enabled diverse applications such as style transferring between images~\cite{gatys2016image, zhu2017unpaired}, generating images based on texts~\cite{zhu2019dm}, etc.

Recent research has highlighted and merged the two fields, enabling the multi-modal generative power of GenAI.
Works like Transformer~\cite{vaswani2017attention} and Diffusion Model~\cite{ho2020denoising, Rombach_2022_CVPR, NEURIPS2022_ec795aea, nichol2022glide} have built the theoretical foundation for the current stage of GenAI, where large models such as Generative Pretrained Transformer (GPT) and its successors~\cite{radford2018improving, radford2019language, keskar2019ctrl, brown2020language}, T5~\cite{raffel2020exploring}, BERT~\cite{devlin2018bert}, and CLIP~\cite{radford2021learning} enable diverse applications with content of higher quality and multiple modalities.

Empowered by the models, an increasing number of software and interactive platforms are being developed to cater to various fields, such as art, design, education, and algorithm development, democratizing access to the creative potential of GenAI. 
In this paper, we provide a comprehensive summary of diverse GenAI applications and highlight some notable open-source platforms in ~\autoref{table:apptable}
. 
These platforms are designed and contributed by researchers, developers, and experts, aiming to make GenAI technology accessible to a broader audience. 
~\autoref{table:apptable} serves as a valuable resource for readers to gain insights into the versatility of GenAI applications and discover open-source platforms that can facilitate their creative pursuits.

\subsection{Scope}
\subsubsection{GenAI vs AI} 
GenAI, as its name suggests, represents a category of AI that goes beyond traditional models by focusing on generating new data rather than solely analyzing or making predictions. 
In our research, we place a particular emphasis on GenAI models that excel at generating fresh content. 
While traditional AI models are designed to perform specific tasks or offer predefined responses based on data patterns and algorithms (i.e. discriminative), GenAI systems possess the unique ability to create novel content (i.e. generative). 
By enabling users to influence the generated content through inputs like prompts, these interactions become more dynamic and creative.

\subsubsection{GenAI Systems}

Among extensive existing research and work on the applications of GenAI systems, we have chosen to narrow the scope of our research to focus exclusively on GenAI systems that are developed using deep generative models and specifically designed for user interactions, because of their overwhelming generative power~\cite{NEURIPS2021_49ad23d1} and rapid improvement in recent years.

Notably, our paper does not encompass the usage of GenAI models where no user interaction exists, i.e. research that focuses on only the model performance and architectures rather than interactions or applications. 
We have also deliberately chosen not to delve into the detailed formulations of GenAI models and their creation process. 
Similarly, we do not discuss the specific methodologies for creating GenAI models, improving their performance, or training and collecting datasets.
While the technical details of creating and deploying GenAI models are undoubtedly essential and relevant in other contexts, our research emphasizes the human perspective such as the utilization and interaction of these systems by users and the impact of GenAI systems on user experiences, creativity, and decision-making.

\subsection{Contributions}

The GenAI has been explored in various other papers from both sides human and GenAI model~\cite{zhou2023vision+}. Some work has conducted study ~\cite{hu2023exploring} and discussion ~\cite{chen2023next} to gain human perspective in using GenAI models. Chen et al. ~\cite{chen2023next} conducted a discussion with researchers and presented a roadmap for future directions from the technical (GenAI models) side aligning with human values and accommodating human intent~\cite{cao2023comprehensive}. Prior work has contributed in survey and review papers in the field of GenAI such as GenAI recent developments~\cite{zhao2023survey}, their technical perspective~\cite{harshvardhan2020comprehensive, zhang2023text}, content generated~\cite{cao2023comprehensive} and application ~\cite{zhou2023vision+,gozalo2023survey}. Some recent work has also proposed design space \cite{weisz2023toward, hu2023exploring, morris2023design} and design guidelines~\cite{liu2022design}. Recently, GenAI and human interactions have been a topic in HCI workshops~\cite{muller2023genaichi, muller2022genaichi,bernstein2018architecting}.       

Building upon prior work, this paper offers the following significant contributions.

Firstly, it presents a comprehensive \textbf{taxonomy of the design space}, considering both the human perspective and the GenAI perspective.
This taxonomy provides a detailed and in-depth view of various dimensions and categories, with a specific focus on human interactions with GenAI systems. 
By examining the design aspects from these two perspectives, the paper sheds light on the dynamic and creative interactions between users and GenAI systems, providing valuable insights into the user-centric nature of these systems.

Secondly, the paper represents a pioneering effort as the \textbf{first comprehensive literature survey} on human-GenAI interaction systems. 
The literature survey serves as a valuable resource for HCI researchers, offering a well-organized and insightful compilation of the current state of human-GenAI interaction systems.
Researchers can draw from this survey to gain a deeper understanding of the design space and the nuances of interactions between users and GenAI systems. 
Moreover, the survey provides a solid foundation for further research and exploration of novel design possibilities in this rapidly evolving area.

Thirdly, we end our paper with discussions over \textbf{directions for future investigations}, helping researchers identify \textbf{unexplored opportunities and challenges} in human-GenAI interactions.
The discussions and insights are derived from the collections of the papers and the high-level summarization we identified.

%% file: texes/2-Methodology.tex
\section{Methodology}

We aim to identify and collect a large representative set of state-of-the-art GenAI models and GenAI systems using systematic search techniques~\cite{hosseini2023towards}. 
We systematically created a relevant corpus using PRISMA~\cite{Pagen71} guideline: (1) Search strategy to explore; (2) Identification of the publication outlets; (3) Evidence Screening; (4) Eligibility; (5) Inclusion.

\subsection{Search Strategy}

We performed two search waves to select the relevant keywords for our survey.
First search waves identify keywords for advancements in GenAI \textbf{models} in the Machine Learning/Deep Learning conferences.
This also includes open-source software built on GenAI models such as ChatGPT. 
This first spectrum of keywords provides models that were further used in developing a methodology to find GenAI systems corpus in the HCI domain.
The two search strategies are discussed in detail below:   

\subsubsection{Methodology to Define GenAI-related Model Keywords}:
The GenAI field is broad and continuous works are being published in general machine learning areas and domain-specific areas such as CV and NLP.
Therefore, we performed a broad search to identify keywords related to GenAI in order to maximize the inclusion of relevant papers.
Focusing on GenAI \textbf{\textit{models}}, we used different expressions and abbreviations of "Generative Artificial intelligence".
The list of keywords used is shown in~\autoref{tab:kw_model}.
We searched the title, abstract, and keywords to get relevant papers related to GenAI.
The search was conducted in the proceedings of prominent machine learning conferences (ICML, ICLR, and NeurIPS), computer vision conferences (CVPR, ICCV, and ECCV), and NLP conferences (ACL, EMNLP, and HLT-NAACL).
To ensure the relevance of the research specifically to Deep GenAI models, we limited our search to papers published 10 years prior to our work.
During the search process, we carefully examined author keywords, abstracts, and titles to extract additional relevant keywords related to GenAI models.
We also added more keywords based on the three authors' knowledge about the recent advancements in GenAI algorithms and software.

\subsubsection{Methodology to Define GenAI-Related Systems Keywords}:
Then, we focus on finding relevant keywords related to GenAI \textbf{\textit{systems that are interacted with}} by the users.
As the focus of the papers should involve Human-GenAI interactions, we used combinations of keywords: "Generative AI models" AND "Humans" OR "Human" OR "Users" OR "User" for our search.
The complete list of keywords is shown in~\autoref{tab:kw_system}.

\subsection{Identification}

We meticulously executed a systematic search strategy using relevant keywords from renowned publication platforms, including ACM Digital Library, IEEE Xplore, MDPI, Springer, and Elsevier.
Employing the OR operator between keywords ensured a comprehensive exploration of the literature on GenAI models and systems.
Additionally, we proactively searched for variations and synonyms of the keywords, encompassing terms such as "GAN," "StyleGAN," "CycleGAN," "Transfer learning," and "ChatBots," and many more to capture diverse facets of GenAI research.
To focus on the most pertinent content, we applied filters to restrict the search to the title, abstract, and authors' keywords of the articles. 
Considering the rapid advancements in GenAI models, we narrowed our search to papers published in the last 10 years because of the advancements in Deep GenAI models.
Moreover, we prioritized open-access papers and those accessible via institutional subscriptions, broadening the availability of our research findings. We also included relevant citations in our corpus from the papers we found from our key strategy.  

\subsection{Screening}
Three authors screened the initial corpus of 12076 papers individually based on the title and abstract. The entire corpus was divided into 3 equal parts and each part was screened by one author. 
The authors read the title and abstract of the papers assigned to them and screened the paper based on the following exclusion and inclusion criteria: 1) We focus on including papers that were only written in English. 2) We included peer-reviewed conference papers such as journals, conferences, and workshop papers. We excluded opinion papers, idea papers, workshop proposals, patents, non-peer-reviewed papers (e.g. arXiv papers), posters, surveys, and literature review papers. 3) We removed false and change-of-context papers that were not focused on Human-GenAI interaction but were present because of keywords. 4) we excluded the papers with no Deep GenAI model papers used. 5) papers were included that focused on human GenAI interactions rather than on GenAI model performance improvement.
If the authors were unsure about the paper, a discussion among all the authors was considered for the final assessment. After the screening process, we kept 797 papers.

\subsection{Eligibility}
To determine the eligibility of papers each author individually read the entire papers.
As the primary interest is in human GenAI interactions, we included additional inclusion and exclusion criteria (as shown in ~\autoref{table:ECIC}) that were also used when reading the entire paper:
1) we removed papers that were focusing on only technical improvements of the GenAI model.
2) we excluded works without human interactions with the GenAI model or system.
3) we included papers where humans were interacting with the AI system, even if the papers were not focusing on the GenAI system or humans.
4) We included study-based papers that involved interaction with GenAI systems.
These criteria along with screening criteria, three authors reviewed the entire corpus individually by going through the entire paper independently and following the selection criteria to exclude out-of-scope papers.
We determined inter-rater agreement among the three authors using Fleiss' Kappa Measure during the eligibility of the final corpus.
The agreement score of  $\kappa$ = 0.79 indicates substantial agreement.
A final discussion was performed to solve any discrepancies.
The overall focus for selection criteria involves papers with only human GenAI system interactions.
Then all three authors discussed each paper in the corpus to finalize a total of 291 papers. 

\begin{table}[htp]
\centering
\caption{Exclusion Criteria and Inclusion Criteria authors followed during the Confirmation process of the eligibility of the papers.}
\label{table:ECIC}
\small\addtolength{\tabcolsep}{5pt}
{
\begin{tabular}{ll}
\hline
\multicolumn{2}{c}{Exclusion Criteria (EC)}                                                                                                                                                                          \\ \hline
 \begin{tabular}[c]{@{}l@{}}EC1 GenAI Model\\ Technical Improvement\end{tabular} & \begin{tabular}[c]{@{}l@{}} Papers that solely focus on improving the performance of GenAI\\ models for the applications\end{tabular}                                                               \\
EC2 No Human Interactions             & \begin{tabular}[c]{@{}l@{}} Papers that solely present applications of GenAI without actual\\ users or humans interacting with the applications.\end{tabular} \\
EC3 Opinion                           & Literature Review, survey, and opinion papers (reserved)                                                                                                             \\
EC4 Change of Context:                & Papers with words/terms from different contexts from GenAI                                                                                              \\
EC5 False Search:                     & Papers with no desired keywords but found by the search engine                                                                                                    \\
EC6 Idea Paper:                       & Short papers, proposals, demos, and position papers (reserved)                                                                                                       \\
EC7 Non-AI Papers:                    & \begin{tabular}[c]{@{}l@{}} Papers with no GenAI models being used (E.g., generating\\ design using simple if-else conditions).\end{tabular}                                                   \\
                        \hline
\multicolumn{2}{c}{Inclusion Criteria (IC)}                                                                                                                                                                          \\ \hline
IC1 Interactions:                     & \begin{tabular}[c]{@{}l@{}} Papers where humans are interacting with the AI system, even if\\ they are not focusing on the systems or humans\end{tabular}       \\
IC2 Extended Abstracts:               & Papers published as extended abstracts                                                                                                                          \\
IC3 Study-based Papers                & Study-based papers that involve interaction with GenAI systems.                                                                                                 \\ \hline
\end{tabular}
}
\end{table}

\subsection{Inclusion}
Our final paper has a total of 291 papers. The collected corpus varies from 2013 to 2024. A larger majority of the papers have been published in ACM (55\%)and IEEE (29\%). \autoref{tab:data} shows the filtering of manuscripts at various stages for the final corpus. 

Despite our systematic methodology for collecting representative papers of the existing work, we acknowledge that the resulting corpus might not contain every piece of the development of GenAI, due to the mass popularity of GenAI in every other domain such as medicine, manufacturing, and others.
Further, there might be some papers that are on the borderline of Exclusion/Inclusion criteria that were not included in the corpus.

Nevertheless, we argue that this corpus is large and comprehensive enough to provide representative subsets of each dimension for the most relevant papers for our taxonomy.
To collectively address our limitations, we will make our coding, dataset, and tagging system open-source and available for researchers to iterate, suggest new categories, expand our current taxonomy, and advance.

\begin{table}[htp]
\caption{Manuscripts filtered at various stages. Id. for Identification, S. for Screening, E. for Eligibility.}

\label{tab:data}
\resizebox{\columnwidth}{!}{%

\begin{tabular}{cccccccccc}
\hline
Process        & Criteria   & ACM  & Springer & IEEE & S.D. & MDPI & T\&F & Others & Total \\ \hline
Identification & Search term         & 1279 & 1048     & 6832 & 1334          & 1164 & 443 & 14 & 12114 \\
Identification & Filtering            & 1265 & 1048     & 6812 & 1332          & 1164 & 441 & 14 & 12076 \\
Screening    & Keyword and abstract & 138  & 35       & 394  & 47            & 13   & 46  & 14 & 797   \\
Eligibility    & Full paper           & 160   & 6        & 84   & 7             & 11   & 9   & 14  & 291   \\ \hline
\end{tabular}%
}
\end{table}

\begin{figure*}[ht]
    \centering
    \includegraphics[width=0.8\textwidth]{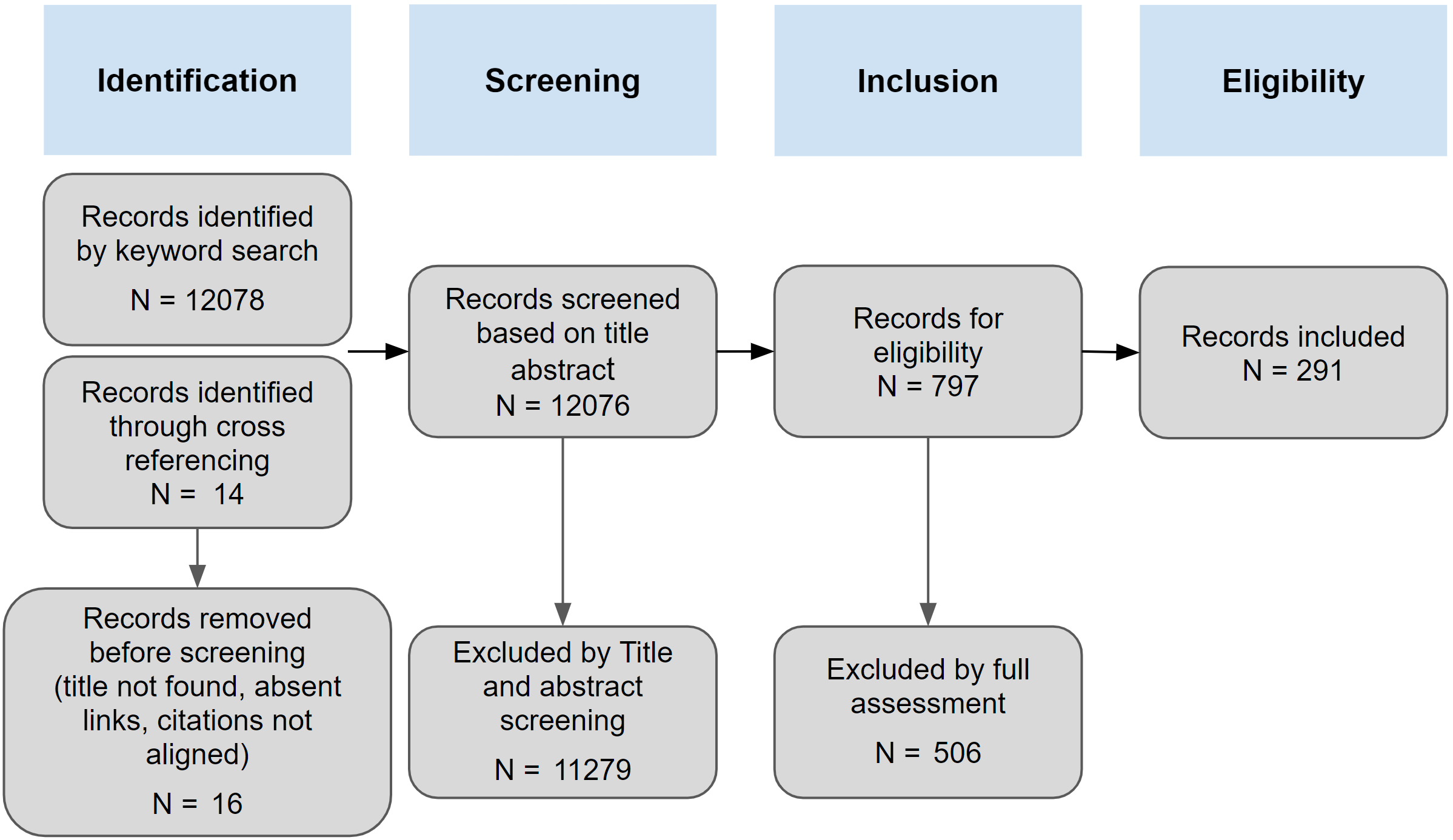}
    \caption{Flowchart for our paper selection process}
    \label{fig:flowchart}
\end{figure*}

\subsection{Analysis}

As discussed in the former sections, we aim to present a comprehensive taxonomy of human-GenAI interaction in order to help future research and development in designing interactive GenAI systems.
Over the collected corpus of literature, we conduct a collaborative analysis to identify the key dimensions of the taxonomy.

The analysis involves multiple stages of review and collaboration among the authors. 
Initially, each of the authors reads and summarizes a small subset ($N=25$) of the papers to establish an approximate taxonomy of components and dimensions.
These preliminary taxonomies are then discussed among all authors to iteratively refine and enhance, by highlighting the overlapping dimensions, adding or subtracting components and categories as necessary, and eventually agreeing upon one revised taxonomy.
Upon revision, three authors individually read the entire list of papers thoroughly to assign them to their respective categories and dimensions.
During this phase, workshop proposals, surveys, and literature reviews are not included in the final categorization but are used as supplementary references to guide our analysis and concretize the design components and categories.  
To ensure consistency and resolve any conflicts, the three authors subsequently engaged in discussions to finalize the tagging of the papers, by iteratively opening up new discussions upon any author identifying a paper that does not properly fit into the existing taxonomy.
At the end of each iteration, authors decide if a new dimension is needed for the taxonomy or agree on the position of the paper in the existing taxonomy.
This collaborative approach helps ensure the accuracy and reliability of the categorization process.

Eventually, our analysis results in a comprehensive taxonomy of dimensions from an HCI design perspective.
The HCI design logic behind this taxonomy suggests that developers and researchers in this field practice User-Centered Design~\cite{abras2004user} by
\begin{enumerate}
    \item orienting the system through the needs of the end-users (\autoref{sec:purpose})
    \item considering the capabilities and functions of the GenAI models (\autoref{sec:feedback})
    \item exploring the possible actions of the users, the corresponding results of the actions, and the mapping between the actions and the results(\autoref{sec:control})
    \item enabling different levels of engagement in the interaction for different tasks (\autoref{sec:engagement})
    \item deciding the application domain the system can be generalized into (\autoref{sec:application}),
    \item and evaluating the system via designated metrics (\autoref{sec:evaluation}).
\end{enumerate}

In the following sections, we discuss various components in the dimensions and the motivations for considering them when designing a system.
Later on, we present the future opportunities and challenges in this domain summarized from our literature review.
In the appendix, we included tables containing all the papers' citations and counts that fall into respective categories and dimensions.



%% file: texes/3-Purpose.tex
\section{Purposes of Using GenAI}
\label{sec:purpose}

The initial step of designing an interactive GenAI system is to determine the end goals of (i.e. purposes of using) the system from the user's perspective.
Characterizing such purposes sets the fundamental notes on the design of the systems and answers the essential questions on what types of tasks the system is to fulfill and what the users of the system expect from it.
To this end, in the very first section, we categorize the purposes of users of GenAI applications.
On a high level, we identify the purposes falling into the following categories: 1) Refine the Outcome, 2) Explore Alternatives, 3) Get Answers for Inquiries, 4) Understand a Subject, 5) Automate Processes, 6) Enhance Experiences, and 7) Augment Sample Data.

\begin{figure*}[ht]
    \centering
    \includegraphics[width=\textwidth]{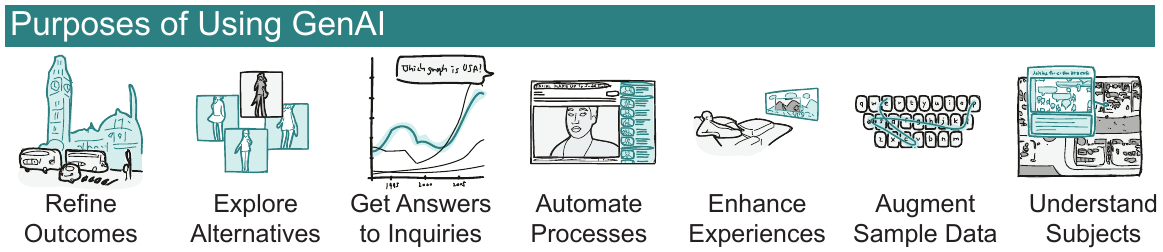}
    \caption{Purposes of Using GenAI depict the users' intention of the interactions and the high-level capabilities of the applications, consisting of Refine Outcomes~\cite{bau2020semantic}, Explore Alternatives~\cite{wan2023gancollage}, Get Answers to Inquiries~\cite{kim2020answering}, Automate Processes~\cite{truong2021automatic}, Enhance Experiences~\cite{shirazi2021supervised}, Augment Sample Data~\cite{chu2023wordgesture}, and Understand~\cite{park2023generative}}
    \label{fig:userexample}
\end{figure*}

\subsubsection*{\textbf{\color{teal}Purpose-1 Refine the Outcome}}
Users possess certain expectations of an instance or outcome to be created.
With a specific objective, users utilize GenAI applications to generate instances to meet their qualitative or quantitative expectations.
Qualitative expectations of the users encompass subjective properties of the instances, such as style of a fashion design~\cite{wu2023styleme}, melody in a piece of music~\cite{louie2020novice, suh2021ai}, plots in a story~\cite{chung2022talebrush}, layout in a web application~\cite{jing2023layout, mozaffari2022ganspiration}, content~\cite{evirgen2022ganzilla, qiao2022initial} or subjects~\cite{de2020brain, chang2023prompt} in an image, etc.
Quantitative expectations depict the objective metrics that the generated instances are to satisfy, such as parametric designs of a 3D model~\cite{koyama2022bo}, efficiency of codes~\cite{liu2023wants}, precise layout of cameras in a VR space~\cite{yoo2021virtual}, etc.

\subsubsection*{\textbf{\color{teal}Purpose-2 Explore Alternatives}}

Users can actively utilize GenAI to generate multiple instances of an idea, building from the abstraction of knowledge learned by the GenAI models.
Users then explore the generated alternatives for ideation, choosing among, or learning the patterns.
For example, GAN-based applications like GANravel~\cite{evirgen2023ganravel} and GANCollage~\cite{wan2023gancollage} enable the users to generate multiple images similar to the input and explore the gallery of images to decide the best design of the images.
GenAI can also passively assist users in their ideation process.
For example, CatAlyst~\cite{arakawa2023catalyst} motivates the users to continue their unfinished presentations by completing part of their work to provide new ideas.

\subsubsection*{\textbf{\color{teal}Purpose-3 Get Answers for Inquiries}}
When faced with a specific challenge or question, users can leverage GenAI to brainstorm potential solutions or avenues of inquiry.
The inquiries put forward can be of single or multiple modalities and from diverse domains.
To design GenAI applications fulfilling this purpose requires considerations over both the delivery of the answers and the capability of providing correct or rational answers.
For instance, users can ask GenAI systems a specific question and have it generate programming codes to solve this question and iterate with the system to optimize the codes~\cite{kazemitabaar2023studying}.
Moreover, GenAI can directly generate the answers in natural language to the problem input by users~\cite{kim2020answering}.

\subsubsection*{\textbf{\color{teal}Purpose-4 Understand Subjects}}
Different from getting answers to specific inquiries, users can utilize the generative powers of models to obtain their understanding of various subjects in a comprehensive manner, without users explicitly deciding the form or content generated to contribute to the understanding.
The understanding is enhanced by GenAI through diverse methodologies such as providing insights, generating examples, or offering new perspectives.
Such understanding can be of the process of the model itself (e.g., GANslider~\cite{dang2022ganslider} utilizes filmstrips of screenshots to illustrate the process of GAN-based transferring of images.), the knowledge of a concept in specific domains (e.g., Liu et al.~\cite{liu2022design} demonstrates how users of text-to-image GenAI can effectively prompt by observing vast amount of text-image-pairs.), the nature or phenomena conveyed by data (e.g., Vis Ex Machina~\cite{zehrung2021vis} generates graphs and charts based on input data from the users in order to help understand the data.).

\subsubsection*{\textbf{\color{teal}Purpose-5 Automate Processes}}
Users can utilize GenAI as an automation tool in a broad spectrum of applications.
GenAI is specifically strong in automating tasks that involve the generation of content.
GenAI can be applied to generating control sequences of robots~\cite{lin2020your, brohan2023can, ren2023leveraging, huang2022inner} in various scenarios based on users' textual or voice commands.
While traditional automation focuses on executing repetitive tasks based on specific instructions, GenAI can also introduce creativity, adaptability, and decision-making into automation processes.
For example, Liventsev et al.~\cite{liventsev2023fully} propose fully autonomous programming with LLMs, where the models are able to rationalize and determine the best practice of coding.

\subsubsection*{\textbf{\color{teal}Purpose-6 Enhance Experiences}}
GenAI, given its ability to generate content and adapt to user input, can significantly enhance user experiences across various platforms and applications.
In general, GenAI is capable of enhancing the experience by improving the quality of the generated content based on diverse metrics.
For instance, GenAI can extend the visual experience of the users~\cite{kimura2018extvision}, make language in articles user-friendly~\cite{strengers2020adhering}, or modify the user input for more efficient communication~\cite{wu2022ai, valencia2023less}
A particular key aspect in GenAI's enhancement of experience is personalization, where GenAI adapts its output based on users' profiles, preferences, or states.
For example, VocabEncounter~\cite{arakawa2022vocabencounter} adapts to the contexts of the users to provide personalized experiences of learning foreign vocabulary.
AdaptiFont~\cite{kadner2021adaptifont} generates adaptive fonts according to users' reading speed to maximize the reading experience.

\subsubsection*{\textbf{\color{teal}Purpose-7 Augment Sample Data}}
GenAI has become a powerful tool for data augmentation, a process used to increase the amount and diversity of data.
AI-generated data can be used as training data to build a new AI model.
For example, Word-Gesture-GAN~\cite{chu2023wordgesture} utilizes GAN to generate synthetic gesture data for training keyboard gesture recognition models.
Deepwriting~\cite{aksan2018deepwriting} uses GAN to generate handwriting data to train style-transfer models.
Enabled by the large corpus of knowledge learned by LLMs and LVMs, research has also investigated the possibility of deploying AI-generated data of various modalities into other research domains.
For instance, Park et al.~\cite{park2022social} and Hamalainein et al.~\cite{hamalainen2023evaluating} experimented with AI-generated for research in social computing and HCI respectively.

%% file: texes/4-Feedback.tex
\section{Feedback from Models to Users}
\label{sec:feedback}

To better understand the essence of human-GenAI interaction, we investigate the field from two major aspects.
The first aspect is the feedback from Models to Users.
In this aspect, we majorly consider the capabilities of the models, the range of modalities of the models, and the methodologies of delivering the output of the models.
These considerations over the models are essential to designing a GenAI system in order to fulfill the purposes aforementioned in~\autoref{sec:purpose}.
Prior work~\cite{yang2020re} in general human-AI interaction has also addressed that the key design challenge stems from understanding AI's capabilities, especially from a user-centered perspective.
To this end, we identified three dimensions to depict the current landscape of feedback techniques of GenAI systems, namely, 1) Output Modalities, 2) Functions of the Models, and 3) Output Synchronization.

\subsubsection*{\textbf{\color{blue}Dimension-1 Output Modalities}}

\begin{figure*}[hbtp]
    \centering
    \includegraphics[width=\textwidth]{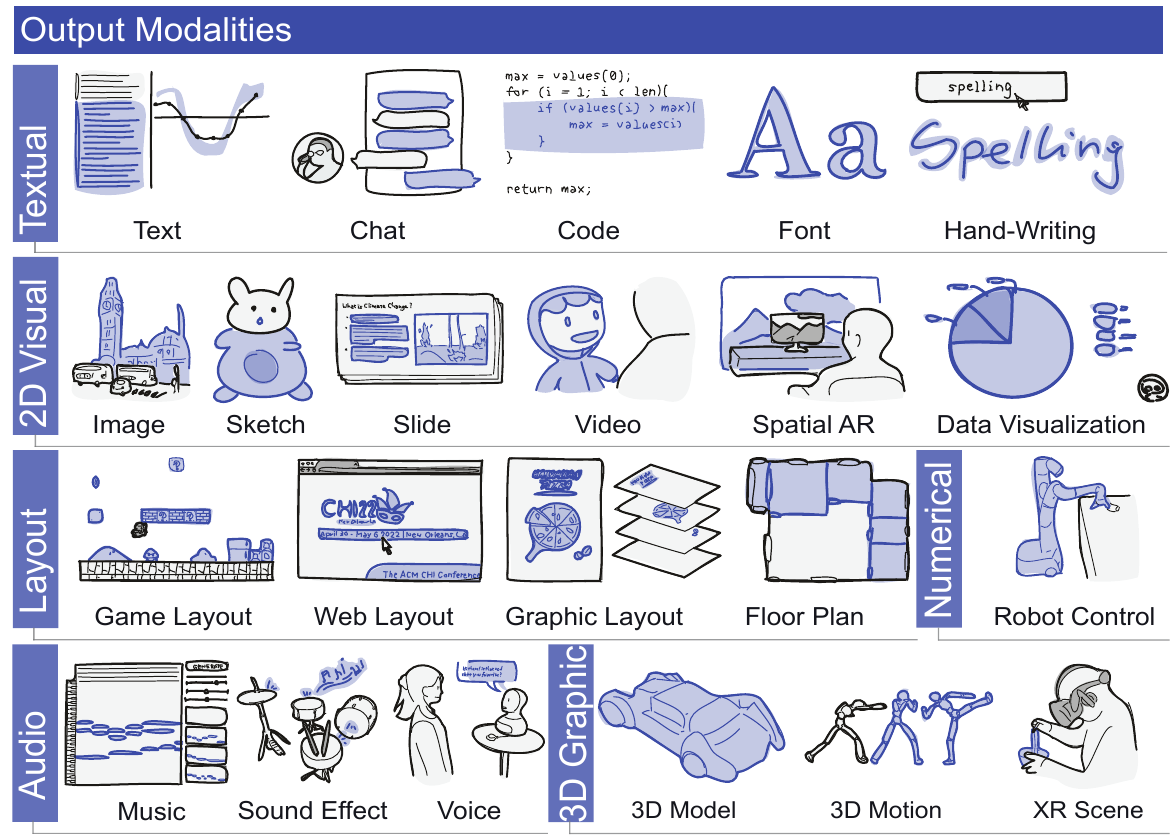}
    \caption{Output Modalities consists of four categories, namely {\color{blue}textual} (text~\cite{chung2022talebrush}, chat~\cite{jo2023understanding}, code~\cite{kazemitabaar2023studying}, font~\cite{kadner2021adaptifont}, and hand-writing~\cite{aksan2018deepwriting}), {\color{blue}2D visual} (image~\cite{bau2020semantic}, sketch~\cite{fan2019collabdraw}, slide~\cite{arakawa2023catalyst}, video~\cite{yoo2021virtual}, spatial AR~\cite{kimura2018extvision}, and visualizations of data~\cite{han2021designing}), {\color{blue}layout} (game layout~\cite{volz2018evolving}), web layout~\cite{kim2022stylette}, graphic layout~\cite{guo2021vinci}, and floor plan~\cite{he2022iplan}), {\color{blue}numerical data} (robot control sequence~\cite{huang2022inner}), {\color{blue}audio} (music~\cite{suh2021ai}, sound effect~\cite{chang2018perceptual}, and voice~\cite{janssens2022cool}), and {\color{blue}3D graphics} (3D model~\cite{liu20233dall}, 3D motion~\cite{xu2021gan}, and XR scene~\cite{nakano2019enchanting})}
    \label{fig:outputmodalities}
\end{figure*}

The output modality of a GenAI model refers to the type or form of data that the model produces.
Generative models can produce a variety of outputs, and the modality is determined by the type of data they are trained on and designed to generate.
In our research, output modality determines the modality of the feedback presented to the users (i.e. they are identical), because we have not identified a case where the output of the system is not presented to the users.

\textit{\color{blue}---Textual}
Textual output encompasses natural language in texts, programming code~\cite{liu2023wants, jiang2022discovering, liventsev2023fully}, the handwriting of texts~\cite{aksan2018deepwriting}, and fonts~\cite{kadner2021adaptifont}.
Specifically, natural language in texts can be chats~\cite{zamfirescu2023herding, huber2018emotional, han2021designing}, descriptions of a problem~\cite{kazemitabaar2023studying}, or pieces of literature~\cite{strengers2020adhering, oh2020understanding,chung2022talebrush}.

\textit{\color{blue}2D Visual}
GenAI models that produce 2D visual outputs are typically trained on large datasets of images to learn the underlying patterns and can create novel images based on their training.
2D visual outputs consist of images~\cite{liu2020ir, chang2023prompt, sanchez2023examining, liu2022design,ross2021evaluating}, sketches~\cite{zhao2020iconate, cheng2020sequential}, videos~\cite{truong2021automatic, liu2023generative}, 2D visualization of data~\cite{zehrung2021vis, lee2022promptiverse}, and spatial AR~\cite{kimura2018extvision, kimura2019deep}.

\textit{\color{blue}3D Graphic}
3D graphic outputs consist of 3D models, 3D motion of various objects, and XR scenes.
For example, Koyama et al.~\cite{koyama2022bo} utilize generated 3D models to provide suggestions during 3D design.
Yoo et al.~\cite{yoo2021virtual} generate VR camera layouts by referring to a clip of a film.

\textit{\color{blue}Audio}
Audio output consists of music~\cite{louie2020novice, suh2021ai, frid2020music}, sound effects~\cite{oh2020understanding}, or natural language voice synthesis~\cite{janssens2022cool}.

\textit{\color{blue}Layout}
GenAI is capable of generating layout information, widely deployed in designing game layouts~\cite{mozaffari2022ganspiration}, web layouts~\cite{wang2023slide4n}, graphic layouts~\cite{jing2023layout, guo2021vinci}, and more domain-specific layout designs (e.g. game layout~\cite{capps2021using, schrum2020interactive, volz2018evolving}).

\textit{\color{blue}Numerical Data}
All modalities of input can be fundamentally regarded as numerical data in the computer science and engineering realm.
In addition to the aforementioned modalities that contain high-level information that can be directly perceived by humans, we identify the numerical data otherwise conveying information and being used as inputs to GenAI e.g., gestural data~\cite{chu2023wordgesture}, control sequence to a robot~\cite{brohan2023can, ren2023leveraging}, and hierarchical representations of concepts~\cite{lee2022promptiverse}.

\subsubsection*{\textbf{\color{blue}Dimension-2 Functions of the Models}}

\begin{figure*}[ht]
    \centering
    \includegraphics[width=\textwidth]{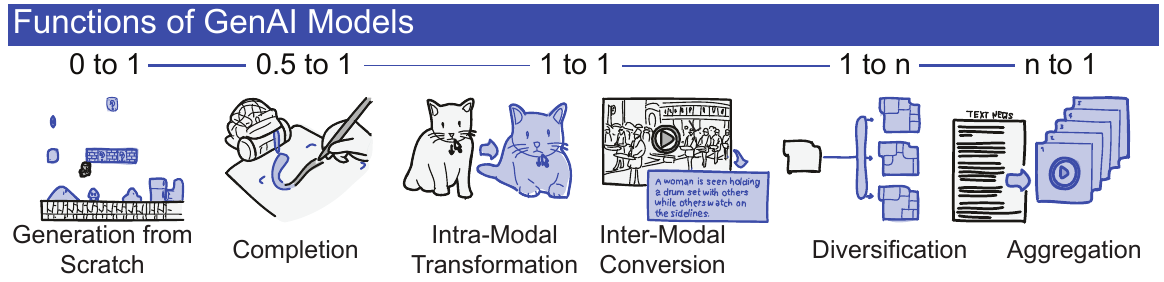}
    \caption{Functions of GenAI Models describe the capabilities of GenAI models, consisting of Generation from Scratch~\cite{volz2018evolving}, Completion~\cite{lin2020your}, Intra-Modal Transformation~\cite{endo2022user}, Inter-Modal Conversion~\cite{wang2021toward}, Diversification~\cite{hu2020graph2plan}, and Aggregation~\cite{oh2020understanding}.}
    \label{fig:functions}
\end{figure*}

As was previously elucidated, the core capability of GenAI is to generate new data samples that are similar in distribution and characteristics to the training data.
Based on this capability, a range of functions of GenAI models are developed.
We categorized the most common six functions in the literature, namely: 1) Generation from Scratch, 2) Completion, 3) Intra-Modal Transformation, 4) Inter-Modal Conversion, 5) Diversification, and 6) Aggregation.
The categorization of the functions of GenAI models aims to bring a clear vision of designing a GenAI system to fulfill the users' needs, utilizing a specific or a combination of functions.

\textit{\color{blue}Generation from Scratch}
GenAI applications are capable of producing entirely new content without specific input.
This could be through utilizing pre-trained patterns, internal algorithms, or some combination of initial states or conditions within the model itself.
The initial generated content is not directly guided by a user's input, and the model operates more autonomously.
With this autonomy, this method of input initialization can benefit (1) GenAI systems with non-expert~\cite{louie2020novice, kadner2021adaptifont, suh2021ai} users who do not possess knowledge of proper input, (2) systems that passively assist the users~\cite{arakawa2022vocabencounter,kadner2021adaptifont}, or (3) systems that help the users with ideation within a specific domain through vast collections of examples~\cite{ko2022we, volz2018evolving, evirgen2023ganravel,padiyath2021desainer}

\textit{\color{blue}Completion}
In some scenarios, GenAI is required to finish an incomplete product from the user.
For example, GenAI can generate auto-completion or suggestions for an ongoing writing task to compose a piece of literature or a story~\cite{jakesch2023co, de2020brain} with designated plots or opinions to inspire or lead the writers.
Moreover, based on what users have input, generated content can be as good as the user input in terms of quality in some cases,~\cite{fan2019collabdraw} or provide a different perspective on the subject~\cite{arakawa2023catalyst}.

\textit{\color{blue}Intra-Modal Transformation}
Intra-modal transformation refers to the function of GenAI to change within the same input modality to generate a different output in the same modality.
Systems that leverage intra-modal transformation typically include modifying the details in the content to meet the users' preferences or experiences (e.g. Strengers et al.~\cite{strengers2020adhering} propose an LLM-based method to modify the article for friendliness to people for minority and De et al.~\cite{de2020brain} enable human portraits editing with brain signals.).
Such transformation usually results in changes in styles~\cite{guo2021vinci, qiao2022initial}, content~\cite{yuan2022wordcraft, qiao2022initial}, or quality~\cite{koyama2022bo, kimura2018extvision, wu2022ai} of the output.

\textit{\color{blue}Inter-Modal Conversion}
Inter-modal conversion refers to the function of GenAI models to convert between different input and output modalities.
This function to convert abstract knowledge or representations among diverse modalities has fostered a promising quantity of possibilities for GenAI applications.
This is because human knowledge and information can now be instantiated to the best modality to either 1) be conveyed efficiently or 2) fit the platforms of the applications.
For example, Cheng et al.~\cite{cheng2020sequential} enable image editing via texts to convert textual descriptions of a design into a visual representation of the design.
Similarly, Yoo et al.~\cite{yoo2021virtual} utilize GenAI to generate VR camera layouts given a reference video, obtaining a unique output for VR applications. 
Moreover, this function of inter-modal conversion has lessened the barriers of expertise requirements in many domains for novices, particularly thanks to its capability to convert ideas and information from intuitive modality, e.g. natural language and sketches, to exclusive modalities, such as programming language, artistic work, or domain-specific designs.
For example, text-to-code applications allow conversion from simple descriptions of tasks in natural language to codes to handle the tasks~\cite{liu2023wants, liventsev2023fully, jiang2022discovering, kazemitabaar2023studying}.
This benefit is also manifested in text-to-image and sketch-to-image applications, where users with no artistic skills can instantiate their intuition or idea, and eventually compose an artistic painting~\cite{chang2023prompt}.

\textit{\color{blue}Diversification}
GenAI possesses the function of diversification, by generating multiple diverse outputs from a single input.
The outputs can be of the same modality (1 to n intra-modal transformation).
In this case, GenAI is capable of generating instances with variations in details.
For example, generating images of the same content but with different viewpoints or features~\cite{zhang2021method, evirgen2023ganravel, wan2023gancollage}, generating longer music given a short clip of melody~\cite{louie2020novice, suh2021ai}, generating textual content such as NPC quests in games~\cite{ashby2023personalized} or (fake) news~\cite{zhou2023synthetic}, or designs for different game layouts~\cite{capps2021using, schrum2020interactive, volz2018evolving}.
Moreover, the outputs can be of different modalities (1 to n inter-modal conversion).
In this case, GenAI is converting the input inter-modally to multiple outputs.
For example, Jing et al.~\cite{jing2023layout} enable the generation of diverse layout designs from a scenario constraint for mobile shopping applications.

\textit{\color{blue}Aggregation}
Finally, GenAI is capable of taking multiple inputs and synthesizing them into a single concise output, which we refer to as aggregation.
GenAI can aggregate inputs of different modalities of inputs to an output of specific modalities.
For instance, PopBlends~\cite{wang2023popblends} blends the concepts from texts and images into a new image, to generate the best representations of an idea.
Huber et al.~\cite{huber2018emotional} make possible the aggregation from texts and images to emotional dialogues.
GenAI can also aggregate inputs to outputs of uniform modalities, focusing on refining the information within.
For example, StyleMe~\cite{wu2023styleme} enables users to merge the outlines and styles from two fashion designs into a new one.
AngleKindling allows journalists to take different angles in writing a journal by summarizing the ideas from the text~\cite{petridis2023anglekindling}.

\subsubsection*{\textbf{\color{blue}Dimension-3 Output Synchronization}}
GenAI systems also utilize different output synchronization strategies.
The synchronization strategies of the GenAI systems investigated follow certain objectives.
Understanding the logic of selecting the synchronization strategy is therefore critical to the system design.
We identify three strategies based on the output timing with respect to the user interaction timing.

\textit{\color{blue}Preliminary}
This category describes a GenAI system with output prior to the user interaction.
The preliminary output strategy is usually utilized in the GenAI systems where users absorb the AI-generated content~\cite{arakawa2022vocabencounter} passively or as is~\cite{jo2023understanding, wang2021soloist}.
Another scenario with preliminary output is the human-GenAI collaboration tasks where GenAI takes the first move to inspire~\cite{lee2022promptiverse} or motivate~\cite{arakawa2023catalyst} the human.

\textit{\color{blue}Real-time}
Real-time output is generated concurrently with human interaction with the GenAI systems.
This strategy benefits the systems with the requirement of immediate responses such as in writing suggestions~\cite{jakesch2023co, valencia2023less, bhat2023interacting} and auto completion~\cite{lehmann2022suggestion}.
Moreover, real-time output is preferred when the systems consist of interaction modalities that tweak the direction, attributes, or details of the generated content.
In these systems, users expect real-time feedback generated when they are interacting.
For example, in GAN-based image generation applications~\cite{dang2022ganslider, xu2021gan, evirgen2023ganravel}, when a user is dragging a slider controlling the direction of the GAN models, the visualization of the generated images is expected to be dynamic and aligned with the slider movement.
This advantage of real-time feedback can be identified through other collaborative applications such as webtoon sketch creation~\cite{ko2022we}, co-writing~\cite{bhat2023interacting}, and programming assistance~\cite{prather2023s, finnie2023my}.

\textit{\color{blue}Delayed}
A delayed output is generated after an explicit mark of the end of the users' interaction, e.g., hitting enter when chatting with a chatbot~\cite{najafian2023people, janssens2022cool} or clicking on a button to input a set of parameters~\cite{guerin2017interactive}.
Delay can be intentionally designed for users to either modify their input or confirm the parameters.
This strategy is common in most human-GenAI interactive applications that require descriptions of human expectations of the output, such as fashion design containing multiple layers~\cite{cheng2020sequential}, artistic image generation considering multiple attributes~\cite{ko2022we}, style merging requiring multiple inputs~\cite{wu2023styleme}, etc.
When there are multiple elements to be considered by the humans in the loop, the delayed output prioritizes users' decision-making on the final output.
Some interaction techniques require delayed output by nature.
For example, interaction with a chatbot requires input and output on a conversational basis which goes one by one.
Nevertheless, the computation cost is a major reason for some applications resorting to delayed output.
Although from a design perspective, real-time output is preferred for the reasons aforementioned, subject to the model size and constraints on the computational power, most image-based GenAI systems resort to delayed output for consistent user interactions.

%% file: texes/5-Control.tex
\section{Control from Users to Models}
\label{sec:control}

The second aspect of the interaction comes from the human side.
We seek the answers to the questions of what and how users can control so that the models arrive at an expected output.
These answers serve as critical human factors in designing the interactions in a GenAI system.
To this end, this section delves into the common ways by which humans can provide feedback to the GenAI system, as well as the categorization of feedback to be provided.
Broadly, we categorize into three categories: \textit{How} users take actions to navigate or adjust GenAI, \textit{what} the objects in GenAI systems are controlled, and the \textit{mediums} to provide feedback.

\subsubsection*{\textbf{\color{magenta}Dimension-1 Methods to Improve the Output}}
In this subsection, we aim to reveal the answers to the essential question -- how do users get the expected generated content from GenAI?
Looking at the question from an HCI perspective, we summarize the high-level interaction design of the GenAI systems in the literature, to identify the methodologies of user interaction to prompt or improve the output towards their expectations.
The design of the interactions follows specific design goals, which are crucial considerations for the HCI design of a GenAI system.

\begin{figure*}[ht]
    \centering
    \includegraphics[width=\textwidth]{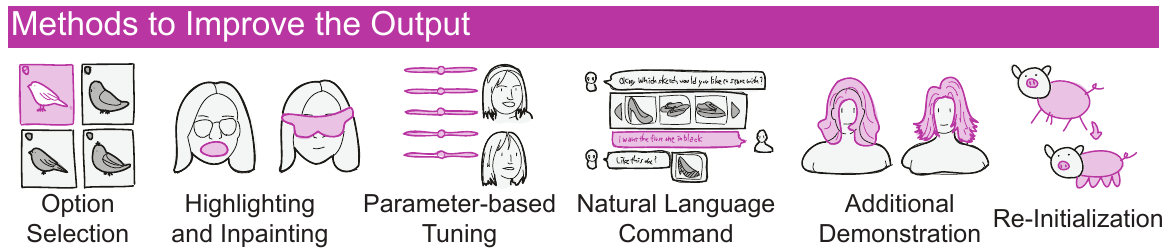}
    \caption{Methods to Improve the Output from the user's perspective include Option Selection~\cite{zhang2021method}, Highlighting and In-painting~\cite{evirgen2023ganravel}, Parameter-based Tuning~\cite{dang2022ganslider}, Natural Language Command~\cite{cheng2020sequential}, Additional Demonstration~\cite{shen2020deepsketchhair}, and Re-Initialization~\cite{fan2019collabdraw}.}
    \label{fig:methods}
\end{figure*}

\textit{\color{magenta}Options Selection}
Users can select their preferred output from a range of options generated by the AI, allowing them to choose the result that best aligns with their needs~\cite{louie2020novice,wan2023gancollage, o2015designscape}. 
Output selection can also serve as a feedback mechanism to train or fine-tune GenAI to get better results in the future.
Additionally, users can also select one of the outputs from GenAI model intermediate layers guiding the direction to the final desired output~\cite{evirgen2022ganzilla}. 
This allows the user to iterate or build on the intermediate output refining them further to achieve the final result.   
 
\textit{\color{magenta}Highlighting and Inpainting}
It allows users to point out specific areas that need modification or replacement in the input such as image ~\cite{evirgen2023ganravel}, text ~\cite{dang2022beyond, fu2023comparing}, or document ~\cite{ma2022ai}. 
Users can highlight or color paint emphasizing particular details, areas, or objects to either add ~\cite{bau2020semantic},erase~\cite{fu2023comparing, bau2020semantic}, modify~\cite{evirgen2022ganzilla, ko2022we} or keep~\cite{lee2022coauthor} specific regions of the input in generating the final output.   
 
\textit{\color{magenta}Parameter based Tuning}
 GenAI system provides users with a unique ability to access and manipulate the intermediate layers to indirectly modify (compared to composing the content artificially) the output~\cite{schrum2020interactive, suh2021ai,ueno2021continuous, dang2022ganslider, koyama2022bo}. 
 It provides users with a granular level of control over the output using slider~\cite{suh2021ai} or numerical input~\cite{ueno2021continuous} to semantically change a generation of output. 
 Parameter tuning is helpful in image input-output target matching~\cite{dang2022ganslider, ross2021evaluating}, controlling the randomness of the generated output such as LLMs~\cite{leinonen2023using}, changing the style of output ~\cite{rafner2021utopian} e.g. graphic design ~\cite{ueno2021continuous}, and editing the content but preserving style ~\cite{rafner2021utopian,aksan2018deepwriting}.       

\textit{\color{magenta}Natural language Commands}
Natural language guidance from humans either text or voice allows them to guide the output from the GenAI system using commands or instructions~\cite{huang2022inner, ren2023leveraging, cheng2020sequential}. 
Users can provide commands sequentially to steer the output generation ~\cite{liu2020ir}. 
These commands are commonly used in Large Language models ~\cite{hamalainen2022neural}, Chatbots~\cite{han2021designing}, and visual design assistant~\cite{cheng2020sequential}. 

\textit{\color{magenta}Additional Demonstration}
Users can further provide additional information to the GenAI system by drawing sketches~\cite{zheng2019content, guerin2017interactive}, outline~\cite{shen2020deepsketchhair}, copy \& paste~\cite{chong2021interactive}, handwriting~\cite{aksan2021generative}, images ~\cite{qiao2022initial} and keywords~\cite{liu2022opal} to narrow the scope of the generated output. 
Users can provide specific preferences ~\cite{zheng2019content}, additional context ~\cite{wu2023styleme, zhao2018compensation}, and set constraints ~\cite{chung2022talebrush} to generate a more relevant and accurate output.  

\textit{\color{magenta}Re-intiliazation}
Users can re-initialize the generation process, with new or some adjustments ~\cite{fan2019collabdraw, lin2020your, twomey2022three} in the input. 
This iterative and adaptable approach allows users to fine-tune content generation effectively ~\cite{oh2018lead}. 
Reinitializing also allows for experimentation to see how different approaches or inputs affect the output ~\cite{linzbach2023decoding}.

\subsubsection*{\textbf{\color{magenta}Dimension-2 Objects to Control}}
In this subsection, we elaborate on the objects inside the GenAI models that are actually controlled by the users during interaction.
We broadly identify four categories, namely, the latent space, the hyper-parameters, the weights/parameters, and the input, which are commonly but not systematically identified objects controlled by the users of an AI system by prior works on general human-AI interaction~\cite{schrum2020interactive, higuchi2021interactive, dudley2018review, amershi2019guidelines}.
As a sub-set of general AI systems, GenAI inherits these identical properties in terms of interactive components.
To provide concrete and novel insights based on prior work by Morris et al.~\cite{morris2023design} that touches upon how GenAI models reach \textit{evolution} of themselves (i.e. how GenAI models improve the output through interactions), we further investigate the distinctions between controlling GenAI and controlling general AI in the following discussion.

\textit{\color{magenta}Latent Space}
Latent spaces are high-dimension representations of the input given by the users. Modifying these high-dimension states allows users a semantically meaningful way to control the style of the output. 
For example, GANravel ~\cite{evirgen2023ganravel} allows users to edit face features such as adding glasses to their eyes or making the person smile keeping the rest of the face the same. 
Such controls of latent space representation in GenAI systems give users the freedom to visualize the direction of output ~\cite{kahng2018gan}, explore for diverse generated outputs~\cite{davis2023fashioning,zhang2021method}, and transform~\cite{mozaffari2022ganspiration} the latent space to get the desired output.  
In most cases, users may not directly manipulate the latent space so they are mapped to UI elements such as clicking a button~\cite{evirgen2023ganravel} or moving sliders~\cite{ko2022we,dang2022ganslider}.  
In addition, latent space allows users to potentially influence the types~\cite{jing2023layout, yuan2022wordcraft}, quantities~\cite{zhang2021method}, and levels of variability~\cite{evirgen2023ganravel} present in the system's outputs.

\textit{\color{magenta}Parameters/Weights}
Parameters/Weights refer to the weights and coefficients that the algorithm extracts (learns) from the data during the training process.
The approach to changing the parameters of a model is addressed as fine-tuning.
Fine-tuning the GenAI system involves re-training the GenAI model either with few or zero-shot learning~\cite{brown2020language, swanson2021story}. 
It allows the system to become adaptive and personalized for each individual user ~\cite{spape2021brain}. 
Fine-tuning GenAI systems makes them personalized to the user by improving task-specific capabilities and domain-specific knowledge of the GenAI system, consequently enabling guided generation of content, in terms of a range of elements in the content, e.g. the subjects or the style of the generated content.

\textit{\color{magenta}Hyper-Parameters}
Hyper-parameters refer to the type of parameters that are second-level tuning parameters that decide the training and inferencing process other than the first-level parameters or model weights~\cite{feurer2019hyperparameter, probst2019tunability}.
Specific designs to interact with these GenAI model hyper-parameters allow users to control the creativity(randomness) in the outputs generated from the GenAI systems~\cite{leinonen2023using}. 
Some GenAI system includes \textit{Temperature} parameter~\cite {valencia2023less, louie2020novice}, frequency penalty ~\cite{lee2022coauthor} and random seed during the development of GenAI system~\cite{leinonen2023using} to control the variability in the generated output. 
For example, Louie et al.~\cite{suh2021ai} use a temperature parameter to generate conventional or surprising music.   
Hyper-parameters can be changed by the end-user using tools that allow changing values using sliders~\cite{louie2020novice} or text editor~\cite{leinonen2023using}. 

\textit{\color{magenta}Input}
Input control provides the unique ability for the user to interact with GenAI without changing the parameters or the hyper-parameters ~\cite{zamfirescu2023herding}.
The quality~\cite{liu2022design} and relevance~\cite{lee2022promptiverse} of the input prompt influence the output generated from GenAI systems. 
Users can provide input prompts to the model to get the desired generated output ~\cite{wu2023styleme}.
Directly giving input prompts from the user to the model often generates out-of-context output. 
To eliminate this users can additionally provide a few examples of the input-output to guide the model in generating output in a way the user wants ~\cite{jiang2022promptmaker, yang2023dual}. 
For example, Jiang et al. ~\cite{jiang2022discovering} used LLMs to support software development.
Users can also develop an automatic method of designing such input prompts instead of manually specifying ~\cite{kim2023metaphorian} to get more relevant output. 
Some of the examples are by suggesting prompts ~\cite{lee2022promptiverse}, constructing prompt templates ~\cite{linzbach2023decoding}, combining multiple prompts primitive ~\cite{jiang2022promptmaker}, reformulating prompts suiting GenAI system~\cite{ma2022ai}, and transforming prompt to different input modality~\cite{chung2022talebrush}.

\subsubsection*{\textbf{\color{magenta}Dimension-3 Mediums of Control}}

\begin{figure*}[ht]
    \centering
    \includegraphics[width=\textwidth]{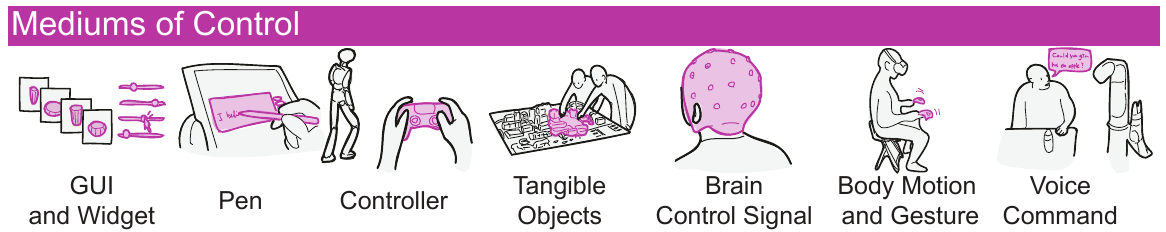}
    \caption{Mediums of Controlling GenAI models include GUI and Widget~\cite{koyama2022bo}, Pen~\cite{aksan2018deepwriting}, Controller~\cite{xu2021gan}, Brain Control Signal~\cite{kangassalo2020neuroadaptive}, Tangible Objects~\cite{noyman2020deepscope}, Body Motion and Gesture~\cite{chang2018perceptual}, and Audio Command~\cite{ahn2022can}.}
    \label{fig:mediums}
\end{figure*}

The interfaces through which users interact with the GenAI models are essential in the HCI design of GenAI systems.
Under a framework of a generic taxonomy of human-computer input mediums, we characterize the interface design considerations for GenAI systems.

\textit{\color{magenta}GUI and Widget}
UI and widgets are the most common design elements in the GenAI system which allow users to understand the displayed information and interact with them, because of the coverage over the most common modalities of data of GenAI systems.
GUIs provide straightforward visualization and mediums of input for textual and imagery data.
Similarly, widgets provide also straightforward adjustments to objects being controlled.
Buttons ~\cite{dang2023choice}, sliders ~\cite{dang2022ganslider}, mouse clicks and drags ~\cite{endo2022user}, and tool pallets ~\cite{zhang2021method} are often used to provide input or any modifications to GenAI system. 
Image editors~\cite{evirgen2022ganzilla} and text editors ~\cite{wu2022promptchainer} are also used to display outputs and provide or edit the inputs. Interfaces are also used in displaying the 3D content \cite{liu20233dall}
Information panels and menus are used to display information that guides users to use the GenAI system ~\cite{suh2021ai}.  
Canvas is also used to provide multiple ~\cite{evirgen2022ganzilla, wan2023gancollage, evirgen2023ganravel} or different modalities~\cite{qiao2022initial} to the user for viewing and selection of the input-output. 

\textit{\color{magenta}Controller}
The controller is a user interface example that allows the user to provide input and can be used to control the GenAI system.
This medium is specialized for the generation of control sequences for robots, game control, or human motion.
For instance, Xu et al.~\cite{xu2021gan} used a game controller to control human motion generation.   

\textit{\color{magenta}Tangible Object}
The user can control GenAI systems by moving objects in the real world~\cite{noyman2020deepscope}. The tangible object includes interactions with physical objects such as static objects\cite{shirazi2021supervised}, dynamic objects\cite{nakano2019enchanting}, and remote objects \cite{duan2019remote}.
Users' interactions with the tangible objects in the 3D space, virtual or real, are usually mapped as the input to the generation of 3D content, with consistency in I/O spatiality.

\textit{\color{magenta}Pen}
For imagery or visual-text-based GenAI systems, having the users draw their ideation or instance is a popular way of controlling the models.
Users can use a physical pen or pencil to draw ~\cite{twomey2022three, lin2020your}, write text~\cite{aksan2021generative, aksan2018deepwriting}, or make sketches ~\cite{ko2022we, zhao2018compensation} on screens~\cite{he2022iplan} and paper ~\cite{twomey2022three}.   

\textit{\color{magenta}Brian Control Signal}
Brian signals are another medium that helps the user control the GenAI system using brain signals. Some works record these signals using electroencephalography(EEG) ~\cite{ kangassalo2020neuroadaptive, spape2021brain, davis2022brain}. Brain responses are directly connected to the internal parameters of GenAI models such as latent space~\cite{de2020brain} for providing implicit feedback. For example, Spape et al.~\cite{spape2021brain} used a brain interface for generating personalized attractive images. 

\textit{\color{magenta}Body Motion and Gesture}
Gesture interaction involves the utilization of physical gestures and movements as a means of engaging with the GenAI system, commonly carrying spatial-temporal information that guides the generation of content.
Gesture movement can be captured by interactive surfaces such as mobiles and tablets~\cite{chu2023wordgesture, chung2021gestural}.
Also, body movements are useful in interacting with GenAI in immersive environments~\cite{chang2018perceptual}.
Face tracking, facial emotion, and expressions are also used in designing natural interactions with GenAI~\cite{jordan2021poseblocks}, which usually carries sentimental or semantical information that guides the generation of content.

\textit{\color{magenta}Audio}
Audio includes both human voice\cite{ahn2022can, janssens2022cool} and music \cite{frid2020music}.
Human voice commands offer a dynamic approach to manipulating and directing the outcomes of GenAI models.
Through vocal prompts, users can effectively steer the generated output~\cite{liu2023wants}, leveraging their spoken instructions to guide the GenAI output.
This innovative interaction method harnesses the potential of natural language and empowers users to shape the GenAI output in a more personalized ~\cite{jo2023understanding} and intuitive manner ~\cite{liu2023wants}.
On the other hand, music or sound effects as the medium of control usually serve as the reference of styles for generated content.

%% file: texes/6-Engagement.tex
\section{Levels of Engagement}
\label{sec:engagement}

With the design human-GenAI interaction investigated from both the human side and the GenAI model side, we then examine the high-level characterization of the design of the interaction as an entity.
In this section, we follow the most commonly used definition of engagement as "\textit{the process by which interactors start, maintain, and end their perceived connections to each other during an interaction.}"~\cite{sidner2003engagement, oertel2020engagement}.
We identify four levels of engagement in human-GenAI interaction, namely, \textit{Passive Engagement}, \textit{Deterministic Engagement}, \textit{Collaborative Engagement}, and \textit{Assistive Engagement}, with an increasing connection between the generated output and the user interaction within the engagement.
Put simply, with a design of passive engagement, users do not substantially contribute to the generation of the content but the models generate the actual output.
On the other hand, in assistive engagement, the models do not substantially contribute but only help in ideation for users to create the content.
A comprehensive metaphor could be writing a story, where the ideation is to come up with the flow and plot of the story and substantial contribution means writing down the story in full-text in any form.

The design of engaging patterns of a GenAI system critically determines the user groups for the particular part of the output users can contribute and the expense of mental load they can afford~\cite{dudley2018review}.
It also determines the application scenarios because the level of engagement limits the choices of interaction mediums.
Finally, the originality or creativity of the content created is also determined by the level of engagement, as pointed out by many prior works~\cite{sarkar2023exploring, mikalonyte2022can, todorov2019game} that the process of the work assigns the credit for generated work by a human-AI team.

\subsubsection*{\textbf{\color{violet}Level-1 Passive Engagement:}}
Passive engagement depicts the systems with which users receive information or content generated by the AI without direct interaction.
In this case, users do not create or contribute to the generated content directly but their profiles or preferences are passively taken by the models to direct the output.
Example system designs with passive engaging interactions fulfill the tasks of immersive news writing~\cite{oh2020understanding} and immersive vision system~\cite{kimura2019deep, kimura2018extvision}, where the users are passively engaged with GenAI and its product without explicit interactions to guide the output.

With only passive input from the users, passively engaging GenAI systems are expected to generate satisfactory products through an end-to-end pipeline.
The design for possible passive input from the users is also tricky and thus challenging due to its implicity and constraints.

\subsubsection*{\textbf{\color{violet}Level-2 Deterministic Engagement:}}
As its name conveys, in a GenAI system where the engagement is deterministic, the outcome is determined by the AI's inherent logic, instructions, or a predetermined set of rules, rather than being shaped by the users' interactions.
Similar to passively engaging systems, deterministic GenAI systems usually consider users' profiles and preferences as part of the input and directly generate content to meet the users' requirements, resulting in limited contribution from the users to the final result.
Yet Deterministic Engagement differs from Passive Engagement with active interactions from the users.
User interactions in this type of engagement are usually just instructions to stop and start the generation.
Examples can be a foreign language dictionary for the users to learn but in AI-decided contexts~\cite{arakawa2022vocabencounter}, AI-generated hierarchical tutorials for the users to follow~\cite{truong2021automatic}, or an adaptive font generator that evaluates the users' performance and autonomously generates the best font for reading~\cite{kadner2021adaptifont}.

An end-to-end pipeline for generation is also expected for deterministic GenAI systems.
In this setup, users do not need to possess knowledge of the domain to contribute to the generation but are supposed to simply notify the systems when the content is needed by them.

\subsubsection*{\textbf{\color{violet}Level-3 Assistive Engagement:}}
Assistive engagement allows the GenAI system to generate content to assist the users in the creation process in the form of suggestions, summarization, or ideation that do not substantially contribute to the final product of the interactions but rather conceptually or abstractly contribute to the creation process.
An example of an assistive system can be an auto-completion assistant in writing~\cite{jakesch2023co}, a contextual provider of suggestions~\cite{valencia2023less}, or an online debugger for an ongoing programming~\cite{prather2023s}.

Assistive engagement requires the systems to possess the capabilities to abstractly understand users' needs and produce high-level assistance with the tasks.
In the possible application scenarios, users of such systems are expected to complete the generation of the final output with the generated high-level assistance from GenAI models.

\subsubsection*{\textbf{\color{violet}Level-4 Collaborative Engagement:}}
Collaborative engagement is the most common design in the current deployment of GenAI systems.
In these systems, GenAI and users work collaboratively on a task, with both sides contributing to the final product.
One major method of these systems is to collaborate through interactive two-way conversation, exchanging information, and user iterating based on the responses.
This method is widely deployed among the systems based on large language models~\cite{mirowski2023co, buschek2021impact, yuan2022wordcraft, jiang2022promptmaker, yang2022ai}, where conversation in natural language dialogue is possible, and some GAN-based applications, where GenAI provides hints or visualizations on the generation direction in response to user queries~\cite{wang2021actfloor,liu2020ir,mozaffari2022ganspiration, evirgen2023ganravel, dang2022ganslider, endo2022user}.
Another method is cooperation, in which the users and the GenAI share the same goal and substantially contribute to the final results in the same format or modality.
Examples of this method can be jointly creating slides for a presentation~\cite{arakawa2023catalyst}, finishing a sketch by GenAI adding details~\cite{fan2019collabdraw}, and composing a piece of melody by both users and generated music~\cite{louie2020novice, suh2021ai}.

Collaboratively engaging GenAI systems require the users to possess the skills and abilities to create the entity or part of content themselves, with GenAI models sharing the same tasks with them.
It also requires the GenAI models to possess a similar set of capabilities as users in the designated application scenarios.

%% file: texes/7-Applications.tex
\section{Application Domains}
\label{sec:application}

\begin{figure*}[ht]
    \centering
    \includegraphics[width=\textwidth]{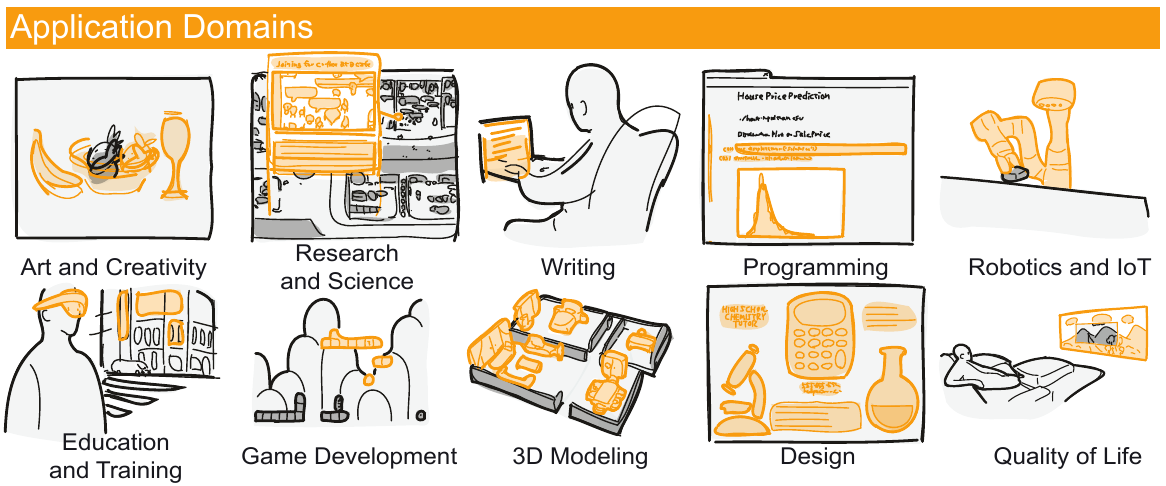}
    \caption{Through our discovery of the literature, the application domains of GenAI include Art and Creativity~\cite{oh2018lead}, Research and Science~\cite{park2023generative}, Writing~\cite{arakawa2022vocabencounter}, Programming~\cite{wang2022documentation}, Robotics and IoT~\cite{ahn2022can}, Education and Training~\cite{arakawa2022vocabencounter}, 3D Modeling~\cite{fu2017adaptive}, Design~\cite{o2015designscape}, and Quality of Life~\cite{shirazi2021supervised}.}
    \label{fig:applications}
\end{figure*}

Through our exploration, we identified a range of diverse application domains of human-GenAI systems.
\autoref{fig:applications} summarizes the categories of domains and lists the related papers correspondingly.
We classified existing works into the following high-level application domains: (1) \textit{\color{orange}Art and Creativity}, (2) \textit{\color{orange}Science and Research}, (3) \textit{\color{orange}
Writing}, (4) \textit{\color{orange}Programming}, (5) \textit{\color{orange}Robotics/IoT}, (6) \textit{\color{orange}Education and Training}, (7) \textit{\color{orange}Game Development} (8) \textit{\color{orange}3D Modeling}, (9) \textit{\color{orange}Design}, and (10) \textit{\color{orange}Quality of Life}.
A detailed list of references in each of the domains above can be found in ~\autoref{tab:applications}.

\textit{\color{orange}Art and Creativity} is the domain where most applications emerge.
The generative power of GenAI has changed the game in the art industry, covering lots of aspects of artistic creation across the disciplines of visual arts, music, literature, and filming.
In general, GenAI can contribute to the processes of ideation, variation, and polishing the artwork.
Such contribution will be further refined by the improvement of the interaction design of human-GenAI.
GenAI also manifests promising potential in the realm of design. 
In \textit{\color{orange}Art} and \textit{\color{orange}Design}, where the visual components are generated by GenAI and then evaluated by human designers, we anticipate further research on the interaction designs in this context from both micro (e.g. efficiency of certain methods for visualizing designs) and macro perspective (e.g. conceptual processes in designs that can be enhanced by GenAI and how).
In our discovery, there are more domains that are less investigated or not investigated best to our knowledge, e.g. \textit{\color{orange}Education and Learning}.
We foresee further exploration and exploitation in these domains based on the insights into specific patterns and methodologies depicted by our taxonomy.

%% file: texes/8-Evaluations.tex
\section{Evaluation Methodologies}
\label{sec:evaluation}

In this section, we report our categorization of evaluation strategies for GenAI systems.
The main categories we identified following the 
classification by Suzuki et al~\cite{suzuki2022augmented}: (1) technical evaluations, (2) evaluation through demonstration, and (3) user evaluations.
Through this section, we aim to provide references for future research on GenAI systems, specifically for deciding the evaluation techniques for future systems.

\subsubsection*{\textbf{Evaluation-1 Technical Evaluation}}
\textit{Technical Evaluation} focuses on the performance of the backend, the algorithm, and the model of a system.
Typical technical evaluation methods on system performance are qualitative assessment of the output~\cite{zhou2023synthetic, leinonen2023comparing, fu2017adaptive}, and quantitative measurement of the output via computing distance (e.g. BLEU for text and FID for images) between the generated and the expected in public datasets~\cite{he2022iplan, davis2022brain, wang2023reprompt, janssens2022cool}.
The evaluations can be conducted on annotated datasets by the researchers themselves and the technical statistics of the datasets are also reported~\cite{aksan2018deepwriting, wang2021soloist}.

\subsubsection*{\textbf{Evaluation-2 Demonstration}}
Evaluations through \textit{demonstrations} assess the system performance under specific conditions.
Common methods consist of generalizability demonstration~\cite{koyama2022bo}, proof of concept demonstration~\cite{strengers2020adhering, lehmann2023mixed, zamfirescu2023herding}, demonstration through an example use case~\cite{yurman2022drawing, chang2023prompt, kimura2018extvision}.

\subsubsection*{\textbf{Evaluation-3 User Evaluation}}
\textit{User evaluation} refers to measuring the performance of a system through user studies, focusing mostly on the effectiveness of the interaction designs in the system, which is hard to technically evaluate through uniform metrics.
Common methods for user evaluation are questionnaires carefully designed to assess how well the design goals of the systems are satisfied~\cite{dang2022ganslider, liu2022design}, qualitative lab studies~\cite{janssens2022cool} for rich insights into design and contextual understanding, quantitative lab studies~\cite{evirgen2022ganzilla} for objective measurement and generalizability, and interviews with both experts~\cite{sun2022investigating, padiyath2021desainer} and novices~\cite{zamfirescu2023johnny, wu2022promptchainer} in the subject matter.

%% file: texes/9-Discussion.tex
\section{Findings}
\begin{figure*}[ht]
    \centering
    \includegraphics[width=\textwidth]{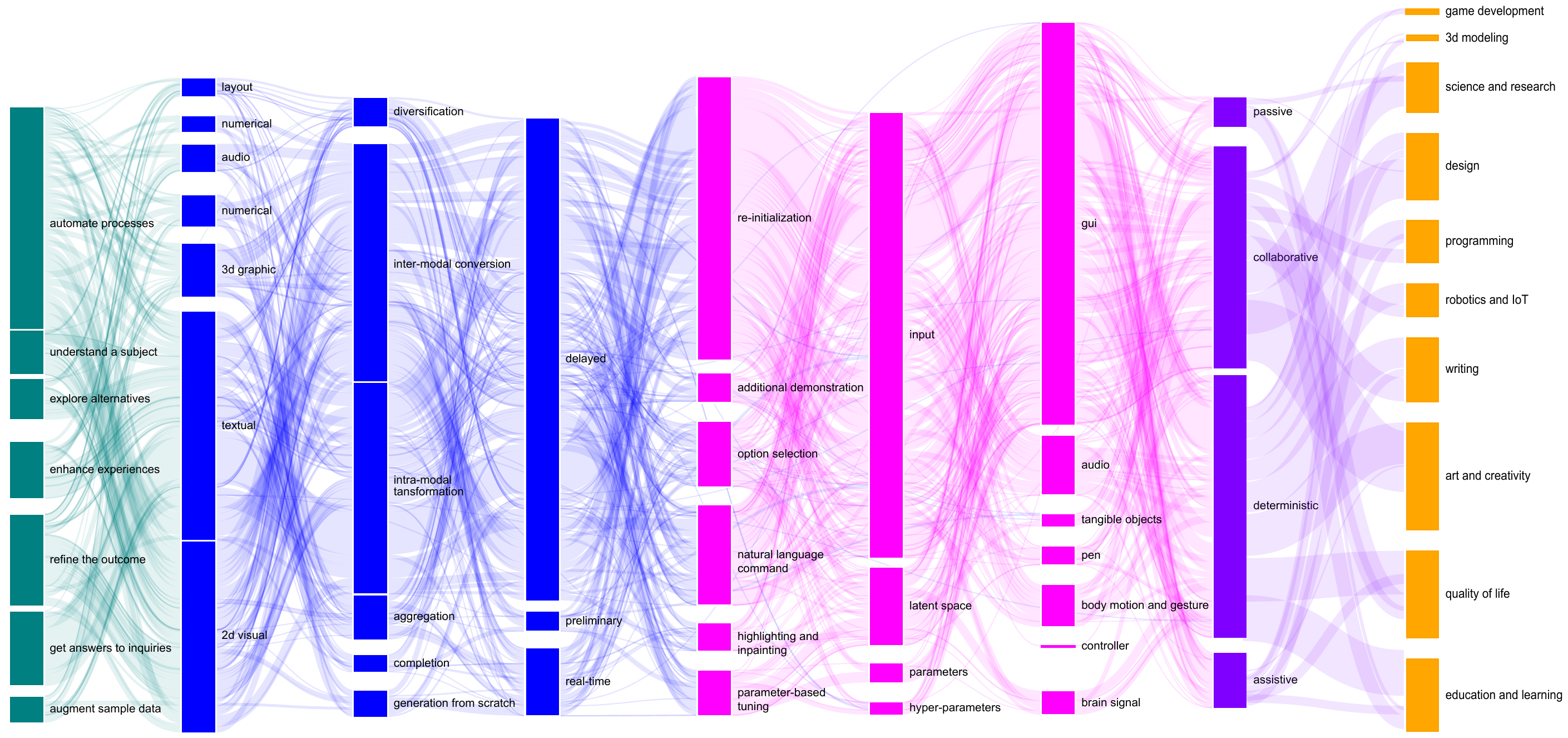}
    \caption{An alluvial diagram of the characteristics from our survey across all dimensions.}
    \label{fig:alluvial}
\end{figure*}

Based on the analysis of our taxonomy, ~\autoref{fig:alluvial} shows a summary of the number of papers for each dimension
In this section, we discuss the standard strategies and gaps that we identify through our analysis of the literature.

\subsubsection*{\textbf{Finding-1 Mediums of Control: Direct but not Intuitive}}
\label{finding1}
Through our literature review, we notice that direct control modalities are more applied (e.g. widgets, controllers, drawings and highlighting, and text, $N=234$) than the intuitive ones (e.g. gestures, brain signals, and voice, $N=49$) to control the output of the GenAI system.
Direct control modalities allow users to modify the attributes of the models or data straightforwardly, while intuitive control modalities require mappings from the users' intuition to the functions of the models.

This inclination highlights that GenAI systems align more with the need to tweak the GenAI models for specific functions directly while overlooking the users' need for intuitive interactions.
For example, using sliders to adjust the weights of the attributes of a GAN model~\cite{evirgen2022ganzilla, wan2023gancollage, evirgen2023ganravel} is direct yet not intuitive, because novice users (who do not know AI) do not possess the technical knowledge to understand the correspondences between the attributes and the outputs.
When users interact with systems with straightforward interactions, they need to build the mapping between their interaction input and the output~\cite{abras2004user}, while with intuitive interactions, researchers have preset this mapping for the users.
Put simply, considering the Gen-AI systems as black boxes with unknown I/O correspondence, intuitive interactions foster smoother learning curves of this I/O correspondence than straightforward interactions.
This observation suggests that intuitive human interactions are not yet the mainstream mediums for controlling the GenAI models. 

\subsubsection*{\textbf{Finding-2 Visualizing the Results rather than the Process}}
We notice that most GenAI systems do not reveal the intermediate layers or output to the users ($N=231$).
This is hard to accomplish from the AI-developing side, given the fact that it is hard for end-users to comprehend the mathematical functions that lie within the intermediate layers of the users.
However, from an HCI perspective, we highlight the necessity of investigating the design space of interacting with the intermediate layers of the GenAI systems, which is a significant missing piece in the current research.
It is important that users understand the intention-action mapping~\cite{abras2004user, norman1988psychology} of using a system, i.e. what consequences result from each of their interactions.
From the designer's perspective, it is required to understand the complexity of the models so that designers can "conceptualize the system’s behaviors in order to choreograph its interactions"~\cite{yang2020re}.
From the user's perspective, users need to interpret outputs and express feedback to the models~\cite{dudley2018review}, which can benefit from a multi-layer design of the system compared to an end-to-end one.
Starting from this consideration, we further discuss the future directions to address this concern in ~\autoref{oppo:designing-interaction}


\subsubsection*{\textbf{Finding-3 the Use of Foundation Models}}
\label{findingFM}
We observed that a major GenAI utilized in the research is Large Language Models (LLMs, a significant subset of Foundation Models, which are defined as models (e.g., BERT, DALL-E, GPT-3) trained on broad data (generally using self-supervision at scale) that can be adapted to a wide range of downstream tasks~\cite{morris2023design, bommasani2021opportunities, kirillov2023segment} in 82 papers). 
Large language models have gained significant popularity due to their unparalleled ability to understand and produce human-like text.
We also observed that most LLM-based applications utilize textual conversation as their interaction modality, implying that the users of these LLM-based systems interact by text input.
This interaction follows the most instinctive patterns for LLMs, which are, after all, models of language.
However, considering LLMs' overwhelming generative power and multi-modal potential, we suggest taking one step back and reconsidering the possible interaction modalities applicable to LLMs.
For example, a voice command can be converted to text to converse~\cite{kim2022stylette}, and vice versa~\cite{janssens2022cool}.
Similarly, images can be summarized by models and translated into text for conversation as well~\cite{wang2021toward}.
Enlightened by this finding, we discuss the future opportunities of research in interaction design in ~\autoref{oppo:fm}

\subsubsection*{\textbf{Finding-4 Missing Discussion over Ethics}}
Through our discovery in the papers, we identify the missing piece of discussion over the ethical problems induced by the widespread application of GenAI.
Out of the corpus of 291 research papers, only 21 papers discuss the potential ethical problems induced or tackled by their systems or studies.
From our previous analysis of the papers, we identified similar patterns in the topics, methodologies, or application domains.
We conclude that the applications of GenAI share similar ethical concerns that are yet to be addressed through further research.
Examples of GenAI ethical problems we have located include GenAI plagiarism~\cite{francke2019potential, DIEN2023108621, noci2023merging}, opinionated bias in GenAI system~\cite{jakesch2023co}, and gender bias in Natural Language generation~\cite{strengers2020adhering}.
We will be detailing the future opportunity of investigating how to tack GenAI ethical problems in \autoref{oppo:ethics}.

%% file: texes/91-Future.tex
\section{Future Opportunities}

\subsubsection*{\textbf{Opportunity-1 Designing and Exploring Interactions with Foundation Models}}
\label{oppo:fm}

\textit{---Various I/O Modalities through Foundation Models}
In our survey, we identified promising usage of Foundation Models, as mentioned in ~\autoref{findingFM}.
While the Foundation Models have enabled diverse applications in domains associated with texts and images, we argue that further research can aim toward \textit{\textbf{more intuitive modalities}}, considering the cross-modality potential shown from both applications we've investigated~\cite{sanchez2023examining, liu20233dall, wang2023reprompt}.
Human conveys information through diverse means in addition to text and image.
For example, the audio of natural language speaking can be converted into text as an approach to converse, a gesture or sign language may contain the information needed for instructing a robot, or a human gaze can guide the foundation models to generate descriptions of an object or an event in sight for educational purpose.
AI and ML community have been widely discussed over the capabilities of Foundation Models learning the abstraction of human knowledge and generating representations of the knowledge in diverse modalities~\cite{morris2023design, bommasani2021opportunities}.
With our proposed taxonomy identifying the design considerations of interaction mediums and modalities of the GenAI models, questions are to be addressed such as (1)\textbf{What are the possible modalities of Foundation Models that can be applied in interactive systems?} For example, there is a limited amount of research that tackles the 3D content generation for AR/VR applications through Foundation Models. (2)\textbf{What are the effects of the conversion of modalities by Foundation Models on the users?} The abstraction of knowledge can be instantiated by Foundation Models across modalities, but what are the effects of the conversion on the users? How much of the information or knowledge is preserved through the generated modalities? How much of it is actually perceived by the users? Answering those questions will significantly guide the direction of future GenAI application development.
(3) The former questions lead to the investigation of another question\textbf{What are the metrics to evaluate the interactions?} Since the popularity and continuing growth of Foundation Models are still novel, there is not yet a set of dedicated metrics to evaluate the quality of the interaction (not just of the model performance or the generated content as proposed in~\cite{muller2023genaichi}) design for them.
Finally, (4) \textbf{What are the general patterns we can conclude from the designs addressing the aforementioned questions?} A general framework for designing interactions with Foundation Models is called for the future development of this promising direction.

\textit{---Diverse Applications through Foundation Models}
Further from above, we argue that more diverse applications of Foundation Models can be introduced by future endeavors.
First of all, through the capability of Foundation Models to handle diverse modalities I/O, \textbf{we anticipate consideration of the formats of data that were unable to be generated by the predecessors of current GenAI models}, which can be specifically used in a certain task.
For example, there can be an application to generate a blueprint of a novel refrigerator (sketch, numerical data, and text as output) given users' routines of menus (text and image as input).
Secondly, \textbf{we suggest that the interaction with Foundation Models (or GenAI in general) should not be constrained to merely collaborative tasks, but can also be applied to tasks with passive or deterministic engagement.}
To be specific, with the strong generative power and capability to consume data in diverse forms, Foundation Models are able to actively understand the environment or context of the users and generate content that is to be passively consumed by the users.
For example, a GenAI-based instructional AR system can scan the vision or environment of the users and detect the elements in the context (e.g. tools, furniture, and appliances), based on which it will predict the intended tasks of the users and generate corresponding AR instructions.
To embrace the promising possibility of diverse applications through Foundation Models, questions remain unanswered \textbf{What are the types of information that can be passively perceived by the users and meanwhile be generated by the Foundation Models (or GenAI in general)? What are the types of contextual, semantic, or environmental information that can be used as the input to the models?}


\subsubsection*{\textbf{Opportunity-2 Designing Interactions for Understanding GenAI}}
\label{oppo:designing-interaction}
Pointed out by many prior works on human-AI interaction, a key consideration in designing an interface for AI-based applications is to enable the users to understand AI, in terms of the capabilities or potential mistakes to make~\cite{amershi2019guidelines}.
Some~\cite{dudley2018review} also argue that the users require the "\textit{ability to retrace steps in the event that recent actions have resulted in an undesired outcome}" and that it is necessary to consider how to reduce the effort of the users to interpret the outputs and express the feedback.
Based on these general discussions over human-AI interaction design, we investigate what open research questions are yet to be addressed specifically about GenAI.

\textit{---Users Understanding the model capabilities}
Different from predictive or discriminative AI models that perform a discrimination behavior, solving classification or regression problems~\cite{gozalo2023chatgpt}, GenAI models generate new media.
This difference creates a divergence between the interactions with predictive AI and those with GenAI.
The capabilities of GenAI to be understood are vaguer and more subjective than those of predictive AI since the latter mostly focuses on or works towards a specific or explicit goal while the former focuses on creativity or implicit tries to meet the users' expectations~\cite{muller2022genaichi}.
The questions remain unanswered:
(1) \textbf{What are the metrics to evaluate how close the generated content is to users?}
With the designated sets of metrics, we can investigate (2) \textbf{How can users themselves learn to interact with the systems to bring the generated content closer to their expectations?}
To address this question, the interfaces or designers need to make clear to the users why the system did what it did~\cite{amershi2019guidelines}.
For example, there are already a massive amount of empirical tutorials, blogs, posts, and even academic papers~\cite{white2023prompt, robinson_how_2023, magalhaes_how_2023} on how to prompt with ChatGPT to direct the output of it.
What are the essential HCI problems behind the empirical studies that can be applied to other GenAI systems?
Answering this question leads to a clearer vision of how interactions should be designed so that the users understand the capabilities of the GenAI models they are interacting with.
(3) \textbf{What are the correspondences between the model parameters and the model capabilities? And which capabilities (and their corresponding parameters) are to be made optional for users to control, considering the particular application domains the systems serve?} Interacting with internal parameters of the GenAI system allows users to explore GenAI model capabilities\cite{dang2022ganslider}.
Only 18 papers out of 291 allow users to control the model parameters.
One of the principles of User-centered Design is that the systems should allow users to guide the behavior of output to align with their preferences~\cite{abras2004user}. 
This involves adjusting certain GenAI model parameters and exploring the full capabilities of GenAI models.
The correspondence between the GenAI model parameters and the actual generated content is different from that between predictive AI parameters and the if-else-rule-based prediction.
A promising direction we see is in the domain of Explainable GenAI, where the correspondence can be studied and applied to the interaction design of actual systems.
It is also crucial for system designers to understand the trade-off between the controllability and the complexity of the GenAI system~\cite{yang2020re}.
Reducing the choices of parameters for the end-users, on the one hand, enhances user experience and increases efficiency, but on the other hand, lessens customization or adaptability of the system.
A balance between the degree of freedom and efficiency of the system has yet to be revealed by future research.

\textit{---Exploring Novel Interactions}
We found limited numbers of dedicated mediums of interaction between the users and the GenAI systems.
With natural interaction such as gesture-based and brain-controlled interfaces, users interact with devices and systems through modalities as intuitive as moving the hands or thinking about a picture, to obtain a desired output.
We foresee the potential novel interactions with such modalities that \textbf{reduce the cognitive offset between user-expected output (resulting from the interactions) and the actual output.}
For example, enabling the usage of brain signals to control the modifications to a generated image waives the cognitive cost of learning the correspondence between traditional GUI and output.

Furthermore, this allows the deployment of GenAI systems in more natural and immersive platforms, particularly in virtual reality (VR) and augmented reality (AR) applications. 
Also, BCIs have the potential to enable interactions without any physical movement, opening up possibilities for users with disabilities and new modes of interaction.
Finally, natural interactions often come with technological difficulties such as user adaptation and personalization, subject to variability in user input and interaction.
For example, people tend to express their feelings in different ways.
Some prefer informal language, while others make ambiguous gestures.
These differences pose challenges for designing an adaptive system that relies on users' expressions as input, say, to generate an image that describes their mood.
The research questions to be addressed in these scenarios are: 1) \textbf{How do we integrate natural interactions with current GenAI models? What can be the connections between the natural interactions and the generation of the content?} 2) \textbf{How do we accurately contextualize and adapt the generated content to users' natural interactions as input?}   


\subsubsection*{\textbf{Opportunity-3 Investigating the Uncertainty of GenAI}}
The uncertainty of an AI system lies in two major aspects.
First, as mentioned in~\cite{yang2020re} the capabilities of the AI system are uncertain to both the HCI designers and the end-users.
Second, the systems are prone to be more uncertain when the users come into the picture as the human factors~\cite{amershi2019guidelines}.
To tackle both aspects of uncertainty, the AI systems are called to be adaptive (as discussed in the last subsection) and explainable respectively.

In the scope of GenAI, the domain of explainability in GenAI is still underexplored, although Explainable AI (XAI) has been discussed for general AI over the years.
The explainability of AI is domain-specific and user-oriented.
Few works like ~\cite{sun2022investigating} investigated the need for explainability for a specific application scenario as well as its impact on the HCI considerations of the systems.


\subsubsection*{\textbf{Opportunity-4 Multi-party Interaction Design}}

A less general yet interesting direction in future discussions of human-GenAI interaction is the multi-party interaction design.
We anticipate a less constrained workflow with multiple users or multiple GenAI models engaging towards the same generative task.
In this workflow, the users and the models can function in the same roles or not.
For example, there can be multiple users brainstorming over a topic while a LLM-based system functions as a summarizer of the discussions from the users.
In addition to the parties, there can be another diffusion-model-based system generating visualization based on the summarization from the LLM-based system.
We propose to investigate the following research question in the potential scenario (1) \textbf{What is the difference in user experience when interacting with multiple GenAI models, when they are of the same and different roles?} (2) \textbf{What are the characterizations of the interactions between the multiple models in the workflow?} More interestingly, (3) \textbf{Can the roles setup be agnostic of whether they are filled by models or humans?} And finally, (4) \textbf{How much can models and humans each contribute to the final content generation?}

\subsubsection*{\textbf{Opportunity-5 Human-GenAI Ethics Discussion}}
\label{oppo:ethics}

As stated in the Discussion and Findings, we identified the missing pieces of discussion over ethical problems induced by GenAI.
The impact of the problems varies across different applications, such as shrinking the job market~\cite{ghosh2022can}, intrusion into copyrights and intellectual property~\cite{noci2023merging}, and generation of illegal content~\cite{al2019privacy, danezis2015privacy}.
We will describe two common problems as examples to open the floor for further discussion for researchers to take into consideration when addressing more foreseen ethical concerns.

\textit{---Credit Assignment between GenAI and Human}
In the GenAI applications where the final output is of market values or artistic attributes, it is still vague and undefined how the credit between GenAI and Human creators should be rigorously distributed, despite the heated discussion over this topic.
The credit assignment dynamic between GenAI systems and human users exemplifies a modern collaboration where innovation is nurtured through a symbiotic relationship.
For example, if an artist creates a painting using a GAN-based system, does he or the GenAI system deserve credit for this artwork?
It would be unrealistic to claim that GenAI should take full credit, for the fact that there would not be art without humans as long as human interactions are the external force fostering this artistic creation.
Yet, one can easily see the flaws and unfairness in giving the human artist the full credit, because, in the creation process, GenAI contributes to the final result, whether in ideation, styling, or any fundamental stroke.
It is also a weak argument that GenAI (or AI in general) is not human and deserves no credit in human work, considering the human efforts in implementing the model and creating the artwork sample training this model.
With all these being said, we propose to take the middle ground that both sides share the credit.
The credit lies in the harmonious exchange: AI offers a canvas, while humans contribute a vivid palette of experiences, cultural nuances, and depth of understanding.
However, a rigorous pattern for credit assignment will not emerge until the following questions are addressed: (1) \textbf{What is the definition of creativity in the context of human-GenAI collaboration?} (2) \textbf{Should the data being used to train GenAI be considered contributing to the generated content?} (3) \textbf{What is the taxonomy of human-GenAI interactions that can help define the contribution of a work?}

\textit{---Inappropriate Use of GenAI}
Generated content can be harmful in many possibilities, such as generating biased or opinionated data for educational content, overlooking the needs of minority groups, generating illegal content that poses threats to society (e.g. rumors), or breaching basic human rights (e.g. identity theft in fake content).
We call for rigorous and clarified rules, regulations, and laws in the domain, which are also considered significant parts of human-GenAI interactions.
Only with clear-defined appropriate applications and usages of GenAI, shall we foster a positive impact of GenAI on the existing human industries and communities.

%% file: texes/92-Conclusion.tex
\section{Conclusion}
In this paper, we present a survey on existing GenAI applications and research, deriving a taxonomy of human-GenAI interactions. 
We synthesize the existing research in this scope and discuss their (1) Purposes of Interacting with GenAI, (2) Feedback from Models to Users, (3) Control from Users to Models, (4) Levels of Engagement, (5) Application Domains, and (6) Evaluation Strategies.
Our research aims to provide an overview of the landscape of the topic of human-GenAI interaction and the common ground of application design.
Further, we discuss future opportunities in this topic, namely, (1) designing interactions for Foundation Models, (2) designing interactions for understanding GenAI, (3) investigating the Uncertainty of GenAI, (4) multi-party interaction design, and (5) ethical discussion on GenAI.
We hope our research will guide and inspire future work on human-GenAI interaction.

%% file: texes/93-Appendix.tex
\appendix
\begin{table}[htp]
\centering
\caption{Appendix Table: Keywords for locating papers about GenAI models}
\label{tab:kw_model}
\small\addtolength{\tabcolsep}{5pt}
{
\begin{tabularx}{\textwidth}{X}
\toprule
"Generative AI" OR "Generative Artificial Intelligence" OR "Deep Generative Modelling" OR “Large language models” OR "LLMs" OR "LLM" OR "Generative Deep Neural Network" OR "Deep Generative Models" OR "Generative Deep Neural Networks" OR "Generative Deep Neural Network" OR “Large Vision Models”\\
\end{tabularx}
}
\end{table}

\begin{table*}[htp]
  \caption{Appendix Table: 
  Keywords used to search for relevant papers of GenAI systems that are interacted with by the users.}
  \label{tab:kw_system}
  \small\addtolength{\tabcolsep}{5pt}
{
\begin{tabular}{l}
                        \toprule
\multicolumn{1}{c}{ACM}\\
                        \hline
\begin{tabularx}{\textwidth}{X}
("Generative AI" OR "Generative Artificial Intelligence" OR "GAN" OR "Generative Adversarial Network" OR "Diffusion Model" OR "language model" OR "Large Language Model "OR "Deep Generative Model" OR "Prompt" OR "LLM" OR "Generative Deep Neural Network" OR "Transformer" OR "GPT" OR "BART" OR "Dalle" OR “CLIP” OR “Codex” OR “Copilot” OR “VAE” OR“Variational Autoencoder” OR “LLama” OR “BLIP” OR “Flamingo” OR “StarCoder” OR “Code T5” OR “Imagen” OR “GLIDE” OR “BERT” OR “Sparrow”) AND("User" OR "Human")
\end{tabularx}\\
                        \toprule
\multicolumn{1}{c}{IEEE}\\
                        \hline
\begin{tabularx}{\textwidth}{X}
("Abstract": "Generative AI" OR "Abstract":"Generative Artificial Intelligence" OR "Abstract":"GAN" OR "Abstract":"Generative Adversarial Network" OR "Abstract":"Diffusion Model" OR "Abstract":"language model" OR "Abstract":"Deep Generative Model" OR "Abstract":"Prompt" OR "Abstract":"LLM" OR "Abstract":"Generative Deep Neural Network" OR "Abstract": "Deep Generative Models" OR "Abstract":"Transformer" OR "Abstract":"GPT" OR "Abstract":"BART" OR "Abstract":"Dalle" OR "Abstract":“CLIP” OR "Abstract":“Codex” OR "Abstract":“Copilot” OR "Abstract":“VAE” OR "Abstract":“Variational Autoencoder” OR "Abstract":“LLama” OR "Abstract":“Flamingo” OR "Abstract":“BLIP” OR "Abstract":“StarCoder” OR "Abstract":“Code T5” OR "Abstract":“Imagen” OR "Abstract":“GLIDE” OR "Abstract":“BERT” OR "Abstract":“Sparrow”) AND ("Abstract":"User" OR "Abstract":"Human")\end{tabularx}\\
                        \toprule
\multicolumn{1}{c}{Springer}\\
\end{tabular}
}
\end{table*}
\begin{table*}[htp]
  \caption{Appendix Table: \color{teal}{Purposes of Using GenAI}}
  \label{tab:purposes}
  \begin{tabularx}{\textwidth}{ccX}
    \toprule
    Category & Count & Citation\\
    \midrule
    \color{teal}\textbf{\textit{Augment Sample Data}} & 17 & Figure~\cite{chu2023wordgesture}:
\cite{du2023automatic, courtial2023deriving, lin2023automatic, wu2023scattershot, kazemzadeh2022bert, hamalainen2022neural, park2022social, fahimi2020generative, chu2023wordgesture, maghoumi2021deepnag, kim2022deep, hamalainen2023evaluating, aksan2018deepwriting, schrum2020interactive, capps2021using, volz2018evolving, ashby2023personalized, zhou2023synthetic} \\
\end{tabularx}
\end{table*}
\begin{table*}[]
  \caption{Appendix Table: \color{blue}Output Modalities}
  \label{tab:outputmodalities}
  \begin{tabularx}{\textwidth}{ccX}
    \toprule
    Category & Count & Citation\\
    \midrule
    \color{blue}\textbf{\textit{2D Visuals}} & 93 &
Figure \cite{chang2023prompt}\cite{fan2019collabdraw}\cite{arakawa2023catalyst}\cite{yoo2021virtual}\cite{kimura2018extvision}\cite{han2021designing}:
\cite{han2021designing, kangassalo2020neuroadaptive,arakawa2023catalyst, davis2022brain, de2020brain, spape2021brain, jordan2021poseblocks, lee2022promptiverse, zehrung2021vis, liu2022opal, wu2023styleme, chang2023prompt, qiao2022initial, liu2022design, wang2023reprompt, koch2019may, chiou2023designing, sanchez2023examining, evirgen2023ganravel, evirgen2022ganzilla, bau2020semantic, zhang2021method, davis2023fashioning, wan2023gancollage, dang2022ganslider, ross2021evaluating, kahng2018gan, chong2021interactive, rafner2021utopian, endo2022user, liu2020ir, wang2023popblends, padiyath2021desainer, zhao2020iconate, cheng2020sequential, truong2021automatic, zhao2018compensation, yurman2022drawing, twomey2022three, fan2019collabdraw, oh2018lead, ko2022we, kimura2018extvision, shirazi2021supervised, kimura2019deep, liu2023generative, ma2022ai,yoo2021virtual, fares2020brain, kavasidis2017brain2image, park2019sanvis, tu2023sdrquerier, bian2021semantic, guo2020attributes, aneja2023clipface, shinagawa2020interactive, lin2021anycost, kahng2020does, pan2023drag, luo2021sketch, abdal2022clip2stylegan, zhou2019interactive, zeng2022aggregated, kaneko2023attempt, luo2021natural, maddigan2023chat2vis, wu2021fashion, sain2021stylemeup, deckers2023infinite, yan2023xcreation, jin2023human, hsu2019dixit, chen2018doodle, bittal2023speech, xu2021textbook, kim202224, lawton2023tool, kanakia2023designing, jin2023gan, fukumoto2018generation, jain2019text, zhang2020sketch, benjamin2023entoptic, wu2023aticvis, derose2020attention, lin2023automatic, siddharth2023blended, li2020screencast, zaramella2023don, arai2023timtoshape, wang2020jean, wang2020jean2, courtial2023deriving}\\
\end{tabularx}
\end{table*}

\begin{table*}[]
  \caption{Appendix Table: \color{blue}Functions of Models}
  \label{tab:functions}
  \begin{tabularx}{\textwidth}{ccX}
    \toprule
    Category & Count & Citation\\
    \midrule
    
\color{blue}\textbf{\textit{Aggregation}}& 33 &
Figure \cite{oh2020understanding}:
\cite{oh2020understanding, liu2022opal, wang2023popblends, wu2023styleme, truong2021automatic, lee2022promptiverse, park2022social, jo2023understanding, arakawa2023catalyst, zheng2019content, zheng2022telling, leinonen2023using, leinonen2023comparing, macneil2023experiences, wang2022documentation, buschek2021impact, kim2019designing, dang2022beyond, lanzi2023chatgpt, mirowski2023co, lee2022coauthor, petridis2023anglekindling, finnie2023my, denny2023conversing, han2021designing, huber2018emotional, ye2018photoreply, maldonado2022path, wu2023aticvis, derose2020attention, kim2021auto, munigala2018persuaide, cai2022pevae}\\

\color{blue}\textbf{\textit{Completion}}& 14 &
Figure \cite{lin2020your}:
\cite{de2020brain, arakawa2023catalyst, twomey2022three, fan2019collabdraw, oh2018lead, lin2020your, prather2023s, dang2023choice, jakesch2023co, lehmann2022suggestion, lehmann2023mixed, sun2022investigating, lelkes2021quiz, guo2022musiac}\\

\color{blue}\textbf{\textit{Diversification}}& 36 &
Figure \cite{hu2020graph2plan}:
\cite{hu2020graph2plan, zhang2021method, wan2023gancollage, koch2019may, davis2023fashioning, evirgen2023ganravel, evirgen2022ganzilla, ko2022we, dang2022ganslider, kahng2018gan, frid2020music, louie2020novice, suh2021ai, chu2023wordgesture, schrum2020interactive, capps2021using, volz2018evolving, o2015designscape, ueno2021continuous, wang2021actfloor, jing2023layout, aksan2018deepwriting, ashby2023personalized, hamalainen2023evaluating, zhou2023synthetic, hamalainen2022neural, valencia2023less, kim2023metaphorian, gero2022sparks, liu2023beyond, kahng2020does, wei2023boosting, umejiaku2023balancing, kwon2019deep, du2023automatic, lin2023automatic}\\
\end{tabularx}
\end{table*}


\begin{table*}[]
  \caption{Appendix Table: \color{blue}Synchronization}
  \label{tab:Synchronization}
  \begin{tabularx}{\textwidth}{ccX}
    \toprule
    Category & Count & Citation\\
    \midrule

\color{blue}\textbf{\textit{Preliminary}}& 14 &
\cite{wang2021soloist, lee2022promptiverse, jo2023understanding, arakawa2023catalyst, lee2019smartphone, shirazi2021supervised, kimura2019deep, arakawa2022vocabencounter, guo2022musiac, arai2023timtoshape, yamaoka2022experience, shirazi2021supervised, kimura2019deep, kularatne2022elderly}\\

\color{blue}\textbf{\textit{Real-time}}& 44 &
\cite{valencia2023less, evirgen2023ganravel, ko2022we, finnie2023my, wang2022documentation, buschek2021impact, lee2022coauthor, prather2023s, dang2023choice, jakesch2023co, lehmann2022suggestion, lehmann2023mixed, sun2022investigating, dang2022ganslider, kangassalo2020neuroadaptive, spape2021brain, naritomi2021ketchup, nakano2019enchanting, xu2021gan, de2020brain, davis2022brain, ross2021evaluating, singh2022hide, calderwood2020novelists, fu2023comparing, bhat2023interacting, zhang2022layout, ye2018photoreply, mertes2023asmrcade, dang2022ganslider, ross2021evaluating, sun2022investigating, nakano2019deeptaste, itadera2021towards, fahimi2020generative, xu2021textbook, gao2019robotic, ito2021estimating, tanaka2023voice, yamazaki2023building, harmouche2020virtual, munigala2018persuaide, cai2022pevae, 10.1145/3610661.3617514}\\
\end{tabularx}
\end{table*}
\begin{table*}[]
  \caption{Appendix Table: \color{magenta}Methods to Improve the Output}
  \label{tab:Improvingoutcome}
  \begin{tabularx}{\textwidth}{ccX}
    \toprule
    Category & Count & Citation\\
    \midrule
\color{magenta}\textbf{\textit{Additional Demonstration}}& 27 &
Figure \cite{shen2020deepsketchhair}:
\cite{chong2021interactive, aksan2021generative, chung2022talebrush, liu2022opal, wu2023styleme, chang2023prompt, qiao2022initial, zhao2018compensation, liu20233dall, duan2019remote, lyu2020real, shen2020deepsketchhair, guerin2017interactive, naritomi2021ketchup, zheng2019content, wang2021actfloor, hu2020graph2plan, liu2023beyond, chung2021gestural, liu2023generative, guo2023pfedprompt, mertes2023asmrcade, zhou2019interactive, yan2023xcreation, lawton2023tool, kim2021auto, lyu2020real}\\

\color{magenta}\textbf{\textit{Highlighting and In-painting}}& 21 &
Figure \cite{evirgen2023ganravel}:
\cite{evirgen2023ganravel, evirgen2022ganzilla, ko2022we, bau2020semantic, lee2022promptiverse, kim2019designing, aksan2018deepwriting, goodman2022lampost, lee2022coauthor, dang2023choice, calderwood2020novelists, fu2023comparing, bhat2023interacting, yuan2022wordcraft, ma2022ai, dang2022beyond, weisz2021perfection, guo2020attributes, zeng2022aggregated, kaneko2023attempt, deckers2023infinite}\\
\end{tabularx}
\end{table*}


\begin{table*}[]
  \caption{Appendix Table: \color{magenta}Objects to Control}
  \label{tab:objects}
  \begin{tabularx}{\textwidth}{ccX}
    \toprule
    Category & Count & Citation\\
    \midrule

\color{magenta}\textbf{\textit{Latent Space}}& 46 &
\cite{jing2023layout, yuan2022wordcraft, zheng2019content, evirgen2023ganravel, evirgen2022ganzilla, ko2022we, bau2020semantic, aksan2018deepwriting, zhang2021method, davis2023fashioning, kangassalo2020neuroadaptive, davis2022brain, mozaffari2022ganspiration, wan2023gancollage, dang2022ganslider, ross2021evaluating, schrum2020interactive, kahng2018gan, chong2021interactive, rafner2021utopian, koyama2022bo, zhou2021interactive, ueno2021continuous, endo2022user, de2020brain, fujita2019perceptual, liu2017interactive, 9533847, mertes2023asmrcade, guo2020attributes, aneja2023clipface, shinagawa2020interactive, lin2021anycost, kahng2020does, pan2023drag, luo2021sketch, abdal2022clip2stylegan, yang2022bodygan, wei2023boosting, cabezon2022capabilities, lu2022preference, deckers2023infinite, kwon2019deep, kim202224, wang2020jean, wang2020jean2}\\

\color{magenta}\textbf{\textit{Hyper-Parameters}}& 13 &
\cite{lee2022coauthor, zamfirescu2023johnny, roemmele2021inspiration, buschek2021impact, louie2020novice, mirowski2023co, prather2023s, suh2021ai, jiang2022promptmaker, leinonen2023using, leinonen2023comparing, valencia2023less, guo2022musiac}\\

\color{magenta}\textbf{\textit{Parameters/Weights}}& 7 &
\cite{brown2020language, spape2021brain, swanson2021story, park2019sanvis, tu2023sdrquerier, bian2021semantic, gao2019robotic}\\
\end{tabularx}
\end{table*}

\begin{table*}[]
  \caption{\color{magenta}Mediums to Control}
  \label{tab:mediums}
  \begin{tabularx}{\textwidth}{ccX}
    \toprule
    Category & Count & Citation\\
    \midrule
\color{magenta}\textbf{\textit{Brain Signal}}& 10 &
Figure \cite{kangassalo2020neuroadaptive}:
\cite{kangassalo2020neuroadaptive, davis2022brain, de2020brain, spape2021brain, fares2020brain, kavasidis2017brain2image, fahimi2020generative, xu2021textbook, wang2020jean, wang2020jean2}\\

\color{magenta}\textbf{\textit{Controller}}& 1 &
Figure \cite{xu2021gan}:
\cite{xu2021gan}\\

\color{magenta}\textbf{\textit{Gesture}}& 14 &
Figure \cite{chang2018perceptual}:
\cite{chung2021gestural, chu2023wordgesture, jordan2021poseblocks, chang2018perceptual, bamani2021robust, jin2023human, adao2023empowering, rajcic2020mirror, mauro2022few, itadera2021towards, dwivedi2022predicting, maldonado2022path, kim2022deep, maghoumi2021deepnag}\\
\end{tabularx}
\end{table*}


\begin{table*}[htbp]
  \caption{Appendix Table: \color{violet}Levels Of Engagement}
  \label{tab:engagement}
  \begin{tabularx}{\textwidth}{ccX}
    \toprule
    Category & Count & Citation\\
\end{tabularx}
\end{table*}

\begin{table*}[]
  \caption{Appendix Table: \color{orange}Application Domains}
  \label{tab:applications}
  \begin{tabularx}{\textwidth}{ccX}
    \toprule
    Category & Count & Citation\\
    \midrule

\color{orange}\textbf{\textit{3D Modeling}}& 9 &
Figure \cite{fu2017adaptive}:
\cite{hu2020graph2plan, lyu2020real, shen2020deepsketchhair, koyama2022bo, guerin2017interactive, yoo2021virtual, fu2017adaptive, liu2017interactive, kim2021auto}\\

\color{orange}\textbf{\textit{Art and Creativity}}& 60 &
Figure \cite{oh2018lead}:
\cite{wang2021soloist, aksan2018deepwriting, zhou2021interactive, louie2020novice, suh2021ai, liu2022opal, twomey2022three, fan2019collabdraw, oh2018lead, lin2020your, frid2020music, chang2023prompt, sanchez2023examining, chang2018perceptual, liu2023generative, ma2022ai, qiao2022initial, zhang2021method, wan2023gancollage, evirgen2022ganzilla, ko2022we, chong2021interactive, rafner2021utopian, bau2020semantic, endo2022user, dang2022ganslider, wang2023reprompt, yurman2022drawing, kangassalo2020neuroadaptive, de2020brain, davis2022brain, spape2021brain, koch2019may, fares2020brain, kavasidis2017brain2image, guo2020attributes, aneja2023clipface, shinagawa2020interactive, lin2021anycost, pan2023drag, luo2021sketch, abdal2022clip2stylegan, lawton2023drawing, zhou2019interactive, zeng2022aggregated, just2023ai, guo2022musiac, deckers2023infinite, yan2023xcreation, chen2018doodle, bittal2023speech, kim202224, lawton2023tool, kanakia2023designing, zhang2020sketch, benjamin2023entoptic, siyao2022bailando, chen2021music, siddharth2023blended, zaramella2023don}\\
\end{tabularx}
\end{table*}


\begin{table*}[]
  \caption{Appendix Table: Evaluation Methodologies}
  \label{tab:evaluation}
  \begin{tabularx}{\textwidth}{ccX}
    \toprule
    Category & Count & Citation\\
    \midrule
\color{black}\textbf{\textit{Technical Evaluation}}& 50 &
\textbf{Quantitative}: 
\cite{aksan2018deepwriting, capps2021using, volz2018evolving, hamalainen2023evaluating, chu2023wordgesture, liventsev2023fully, duan2019remote, ren2023leveraging, ahn2022can, huang2022inner, janssens2022cool, lee2019smartphone, shirazi2021supervised, kimura2019deep, hu2020graph2plan, lehmann2022suggestion, ueno2021continuous, ding2023fluid, lee2022coauthor, gero2022sparks, kim2023metaphorian, wang2023popblends, he2022iplan, zhou2021interactive, shen2020deepsketchhair, guerin2017interactive, sanchez2023examining, bau2020semantic, chong2021interactive, endo2022user, liu2020ir, mozaffari2022ganspiration, wang2021actfloor, cheng2020sequential, chung2022talebrush, kangassalo2020neuroadaptive, davis2022brain, de2020brain, spape2021brain, wang2023reprompt, jing2023layout, lyu2020real, wang2021soloist, truong2021automatic, huber2018emotional, ma2022ai, linzbach2023decoding, chung2021gestural, yang2023dual, guo2023pfedprompt}\\
\end{tabularx}
\end{table*}
\begin{table*}[htp]
\caption{State-of-the-Art Commonly Used GenAI Applications}
\label{table:apptable}
\resizebox{1\columnwidth}{!}
{
\begin{tabular}{lllllll}

\cline{1-5}
\textbf{Input} & \textbf{Output} & \textbf{Application} & \textbf{Model}            & \textbf{Link}                                                                                                            &  &  \\
\cline{1-5}
Text  & Text   & Chatbot     & LLAMA            & {\color[HTML]{0000EE} {\url {https://github.com/facebookresearch/llama}}}                                          &  &  \\
Text  & Text   & Chatbot     & ChatGPT          & {\color[HTML]{0000EE} {\url https://github.com/lencx/ChatGPT}}                                                   &  &  \\
Text  & Text   & Chatbot     & Sparrow          & {\color[HTML]{0000EE} {\url https://arxiv.org/pdf/2209.14375.pdf}}                                               &  &  \\
Text  & Text   & Chatbot     & BART             & {\color[HTML]{0000EE} {\url https://github.com/huggingface/transformers/tree/main/src/transformers/models/bart}} &  &  \\
Text  & Image  & Art         & Dalle -1         & {\color[HTML]{0000EE} {\url https://github.com/lucidrains/DALLE-pytorch}}                                        &  &  \\
Text  & Image  & Art         & Dalle- 2         & {\color[HTML]{0000EE} {\url https://github.com/lucidrains/DALLE2-pytorch}}                                       &  &  \\
Text  & Image  & Art         & CLIP             & {\color[HTML]{0000EE} {\url https://github.com/openai/CLIP}}                                                     &  &  \\
\end{tabular}
}
\end{table*}